\documentclass[twocolumn]{aa}
\usepackage{epsfig}
\def\simlt{\ \raise -2.truept\hbox{\rlap{\hbox{$\sim$}}\raise5.truept   %
\hbox{$<$}\ }}
\def\simgt{\ \raise -2.truept\hbox{\rlap{\hbox{$\sim$}}\raise5.truept   %
\hbox{$>$}\ }}                                                          %

\def\be{\begin{equation}}
\def\ee{\end{equation}}
\def\newline{\hfil\break}

\def\la{\mathrel{\hbox{\rlap{\hbox{\lower4pt\hbox{$\sim$}}}\hbox{$<$}}}}
\def\ga{\mathrel{\hbox{\rlap{\hbox{\lower4pt\hbox{$\sim$}}}\hbox{$>$}}}}

\begin{document}

\title{Structure and evolution of magnetized clusters:\\ entropy profiles, $S-T$ and $L_X-T$
relations}

   \author{S. Colafrancesco
   and F. Giordano
   }

   \offprints{S. Colafrancesco}

\institute{   INAF - Osservatorio Astronomico di Roma,
              via Frascati 33, I-00040 Monteporzio, Italy.\\
              Email: cola@mporzio.astro.it
             }

\date{Received 11 April 2006/ Accepted 29 December 2006}

\authorrunning {S. Colafrancesco \& F. Giordano}

\titlerunning {Structure and evolution of magnetized clusters}

\abstract{We study the impact of an intracluster magnetic field on the main structural
properties of clusters and groups of galaxies: the radial density and entropy profiles,
the entropy -- temperature relation and the X-ray luminosity -- temperature relation for
groups and clusters of galaxies.
To this aim, we develop a description of the intra-cluster gas based on the Hydrostatic
Equilibrium condition and on the Magnetic Virial Theorem in the presence of a radial
distribution of the magnetic field $B(r) = B_* (\rho_g(r))^{\alpha}$, with $\alpha
\approx 0.9$, as the one indicated by observations and numerical simulation.
Our analysis shows that such a description is able to provide, at once, a possible
explanation of three problematic aspects of the cluster structure: i) the flattening of
the entropy profile in the cluster center
ii) the flatness of the $S-T$ relation
iii) the increasing steepening of the $L_X-T$ relation from the cluster scale towards the
group scales.
The available entropy and X-ray luminosity data indicate that an increase of the magnetic
field $B_* \sim T^{0.5 \pm 0.1}$ is required to reproduce at the same time both the $S-T$
and the $L_X-T$ relations.
It follows that a consistent description of the magnetized ICM can provide a simple
explanation of several (or of all) of these still open problems, and thus weakens the
need for the inclusion of other non-gravitational effects which have been proposed so far
for the explanation of some of these features.
This (initial, but not conclusive) analysis can be regarded as a starting point for a
more refined analytical exploration of the physics of the magnetized intra-cluster
medium, and it provides testable predictions that can be proven or disproven with the
next coming sensitive observations of groups and clusters in the X-ray band and in the
radio frequency band.

}

 \maketitle


\section{Introduction}
 \label{intro}

It has been well known since the early 1990s that the intra-cluster (IC) gas properties
deviate from the self-similar predictions of the standard CDM model. The first and most
noticeable piece of evidence is provided by the fact that the X-ray luminosity ($L_X$) --
temperature ($T$) relation is steeper than expected from a standard $\Lambda$CDM model
(see, e.g., Arnaud \& Evrard 1999, Mushotzky 2003, Arnaud 2005 for a review). This
evidence indicated the need for a more detailed look at the IC gas properties. As a
result, the idea that non-gravitational processes can play indeed a role in determining
the cluster structure and evolution has been widely accepted (Evrard \& Henry 1991,
Ponman et al. 2003, see Arnaud 2005 for a review). The picture obtained by the
high-sensitivity, high-throughput X-ray telescopes (Chandra, XMM-Newton) provided a
wealth of precise information on several aspects of the structure and evolution of
clusters and groups of galaxies.\\
The local relations of various physical quantities (i.e., temperature, entropy, X-ray
luminosity, mass) to the temperature of the IC gas (i.e., a measure of their potential
wells and of their equilibrium stage) show that the global IC gas properties like the
X-ray luminosity $L_X$ (Markevitch 1998, Arnaud \& Evrard 1999), the IC gas mass
$M_{gas}$ (e.g., Mohr et al. 1999) or the gas mass fraction $f_g \equiv M_{gas}/M$ scale
with the IC gas temperature $T$ differently than expected from pure gravitational
self-similar evolution of these systems (see Arnaud 2005 for a review). It has been
suggested that the departures of the IC gas scaling laws from the standard self-similar
model are likely due to a non-standard scaling of the mean gas density with temperature
and not to a break of the self-similarity (see Arnaud 2005 for a general discussion).
The gas entropy, defined as $S = k_B T_g/\rho_g^{2/3}$ and related to the true
thermodynamic entropy via a logarithm and an additive constant, is considered in this
context as a fundamental quantity describing the IC gas and therefore can greatly help to
understand the thermodynamic history of the IC gas (e.g., Voit et al. 2002, Voit 2005).
In the standard CDM self-similar scenario, the entropy should scale simply as $S \propto
T$ at any scaled radius. However, it is known that the entropy measured at a radius $r=
0.1 r_{200}$ exceeds the value predicted by the pure gravitation-based model (Pratt \&
Arnaud 2003, Pratt et al. 2006, Donahue et al. 2006, Piffaretti et al. 2005), an effect
that is especially strong for low-$T$ clusters (Ponman et al. 1999, Ponman et al. 2003).
Various non-gravitational processes have been proposed to explain this entropy excess,
such as heating (from SNs and/or AGNs) before or after the cluster collapse, or radiative
cooling (see, e.g., Voit 2005 for a review), but none of these is able to provide a
conclusive explanation of this problem.\\
In this framework, recent XMM data (e.g., Pratt \& Arnaud 2003, Pratt et al. 2006) show a
remarkable self-similarity in the entropy radial profiles for clusters with $k_B T \simgt
2$ keV, a result which is also consistent with a stacking analysis of ROSAT data (Ponman
et al. 2003). The self-similarity of the radial shape of cluster entropy profiles proves
to be a strong constraint for cluster models. For instance, simple pre-heating models,
which predict large isentropic cores, should be ruled out (Arnaud 2005). Furthermore, the
slope of the entropy profiles seems to be slightly shallower than predicted by
shock-heating models (Arnaud 2005). A complex interplay of gravitational, cooling and
galaxy feedback mechanisms is usually considered in the attempt to provide a consistent
picture (see, e.g., Voit 2005 for a review).
It is clear that an unambiguous description of the structure and evolution of groups and
clusters can be obtained only by studying in detail the internal structure of these
systems (see, e.g., Arnaud 2005).

In this context, there has been no attempt, so far, to incorporate a detailed description
of the effect of magnetic fields on the IC gas distribution, irrespective of the fact
that there are several unambiguous and independent pieces of evidence for the presence of
magnetic fields at the $\simgt \mu$G level, especially in the inner regions of galaxy
clusters and groups (see, e.g., Carilli \& Taylor 2002, Govoni \& Feretti 2005) as well
as several numerical (see, e.g., Dolag et al. 2001) and theoretical motivations (see,
e.g., Colafrancesco et al. 2005, see also Giovannini 2004 for a review) for its presence
and extended spatial distribution.\\
In this paper we will explore the role of the magnetic field $B$ in determining the
structural properties of virialized groups and clusters. We will first show in Sect. 2
the effect of the magnetic field in determining the properties of the IC gas in
hydrostatic equilibrium (HE) and their thermal structure from the Magnetic Virial Theorem
(MVT). We will discuss in Sect.3 the entropy profiles of magnetized clusters. In Sect. 4
we will discuss the cluster basic $M-T$ relation and we will show in Sect. 5 that the
inclusion of the B-field provides a flattening of the $S-T$ relation, by acting on the IC
gas thermal and density properties. We will then apply our considerations to the $L_X-T$
relation in Sect. 6 and show that the inclusion of the B-field that reproduce the $S-T$
relation also recovers the observed steepening of the $L_X-T$ relation, from rich cluster
to poor groups. A systematic exploration of the role of the IC gas boundary conditions on
the density, entropy and X-ray luminosity scaling with the cluster temperature will be
provided in Sect. 7. We will finally discuss our results in Sect. 8 in the context of the
existing experimental and theoretical framework, and we will present our basic
conclusions.
The relevant physical quantities are calculated using $H_0 = 71$ km s$^{-1}$ Mpc$^{-1}$
and a flat, vacuum-dominated CDM  ($\Omega_{\rm m} = 0.3, \Omega_{\Lambda}=0.7$)
cosmological model.


\section{The structure of magnetized clusters}

We derive in this Section the relevant quantities that describe the cluster structure in
the presence of a magnetic field. These are the density and the temperature of the IC
gas, both depending, in general, on the magnetic field.

\subsection{Hydrostatic equilibrium of the ICM in the presence of magnetic field}
 \label{sect.he}

The gravitational potential of a galaxy cluster is determined mainly from the DM
distribution which contributes more that $\sim 90 \%$ of its total mass. We assume here,
for simplicity, that the DM density profile is described by the NFW profile (Navarro,
Frenk \& White 1997)
 \be
\rho_{dm}=\rho_s y_{dm}(x)=\rho_s \frac{1}{x(1+x)^2} \, ,
 \label{eq.nfw}
 \ee
where $x \equiv r/r_s$, $r_s$ is the scale radius, $\rho_s$ is a normalization factor and
$y_{dm}(x)$ describes the radial density profile of the DM in terms of the adimensional
radius $x$. The scale radius $r_s$ is linked to the virial radius $r_{vir}$ by means of
the concentration parameter $c \equiv r_{vir}/r_s$, where
 \be
r_{vir} \equiv [M_{vir}/(4 \pi \Delta_c(z) \rho_c(z) / 3]^{1/3} \, ,
 \label{eq.rvir}
 \ee
and $\rho_c(z)$ is the cosmological closure density at redshift $z$. The quantity
$\Delta_c(z)$ is the average density of the DM halo with total mass $M_{vir} \equiv
M(\leq r_{vir})$ in units of the critical density at redshift $z=0$ and takes the value
$\Delta_c \sim 100$ in the reference cosmological model we consider here ($\Omega_m =
0.3$, $\Omega_{\Lambda}=0.7$). We adopt here the predicted value of $\Delta_c(z)$ from
Eke et al. (1996). The concentration parameter $c$ scales mildly with the cluster mass as
 \be
 c = 6 \bigg({M_{vir} \over 10^{14} h^{-1} M_{\odot}} \bigg)^{-0.2}
 \label{eq.c}
 \ee
(see Seljak 2000, Colafrancesco et al. 2006).
Given the previous DM halo structure, the gas density $\rho_g$ and its temperature $T$
can be obtained from the  hydrostatic equilibrium (HE) condition once the equation of
state for the gas is specified. The three cases usually considered are: i) the gas
follows the DM, $\rho_g \propto \rho_{dm}$ (see, e.g., Wu and Xue 2002); ii) the
isothermal case $T_g = $ const (see, e.g., Komatsu and Seljak 2001); iii) the polytropic
case, $T_g \propto \rho_g^{\gamma -1}$ (see, e.g., Komatsu and Seljak 2001, Zhang 2004).
We concentrate here on the isothermal case and we improve it by including the effects of
the magnetic field.

The condition of hydrostatic equilibrium (HE) in the presence of a magnetic field $B(r)$
can be written as
 \be
\frac{dP_g}{dr}+\frac{dP_B}{dr}=-\frac{GM(\leq r)}{r^2}\rho_g,
 \label{terB}
 \ee
where $\rho_g \equiv \rho_g(r,B)$ is the density of the thermal diffuse IC gas and $P_g
\equiv P_g(r,B)$ is its pressure, with both quantities depending in general on the
magnetic field $B$. The magnetic field pressure, $P_B\propto B^2(r)$, adds to the left
hand term of eq.(\ref{terB}).  The magnetic pressure determines, hence, the final
distribution of the IC gas density $\rho_g(r,B)$ since it tends to counterbalance part of
the gravitational pull of the cluster thus prohibiting the gas from a further infalling
that results in a less concentrated gas core.\\
We assume, in the following, a parametric form of the magnetic field radial distribution
 \be
B(r)=B_*\left(\frac{\rho_g(r)}{10^4\bar{\rho}_g(z=0)}\right)^\alpha,
 \label{magn0}
 \ee
where $\rho_g(r)$ is the solution of the HE equation (see Appendix \ref{appendix}),
$\bar{\rho}_g(z=0)$ is the average cosmological density of the gas at $z=0$, $B_*$ is
measured in $\mu$G and $\alpha$ is a parameter $\geq 2/3$ (see, e.g., Zhang 2004, Carilli
\& Taylor 2002, Dolag et al. 2001). We assume in our calculations here a value $\alpha =
0.9$ as indicated by numerical simulations (e.g., Dolag et al. 2001).

In the Appendix \ref{appendix} we have derived  the solution of the HE equation in the
presence of a magnetic field $B(r)$ for two cases of the equation of state of the IC gas:
i) the isothermal case and ii) the polytropic case.\\
We discuss here specifically the IC gas distribution in the case of a isothermal
condition in which  the Magnetic Virial Theorem (MVT) derived for galaxy clusters by
Colafrancesco \& Giordano (2006a) also holds. We will address the polytropic case and
more general cases with a radial dependence of the cluster temperature elsewhere
(Colafrancesco \& Giordano 2006b, in preparation).

\begin{figure}[!h]
\begin{center}
\vbox{
 \epsfig{file=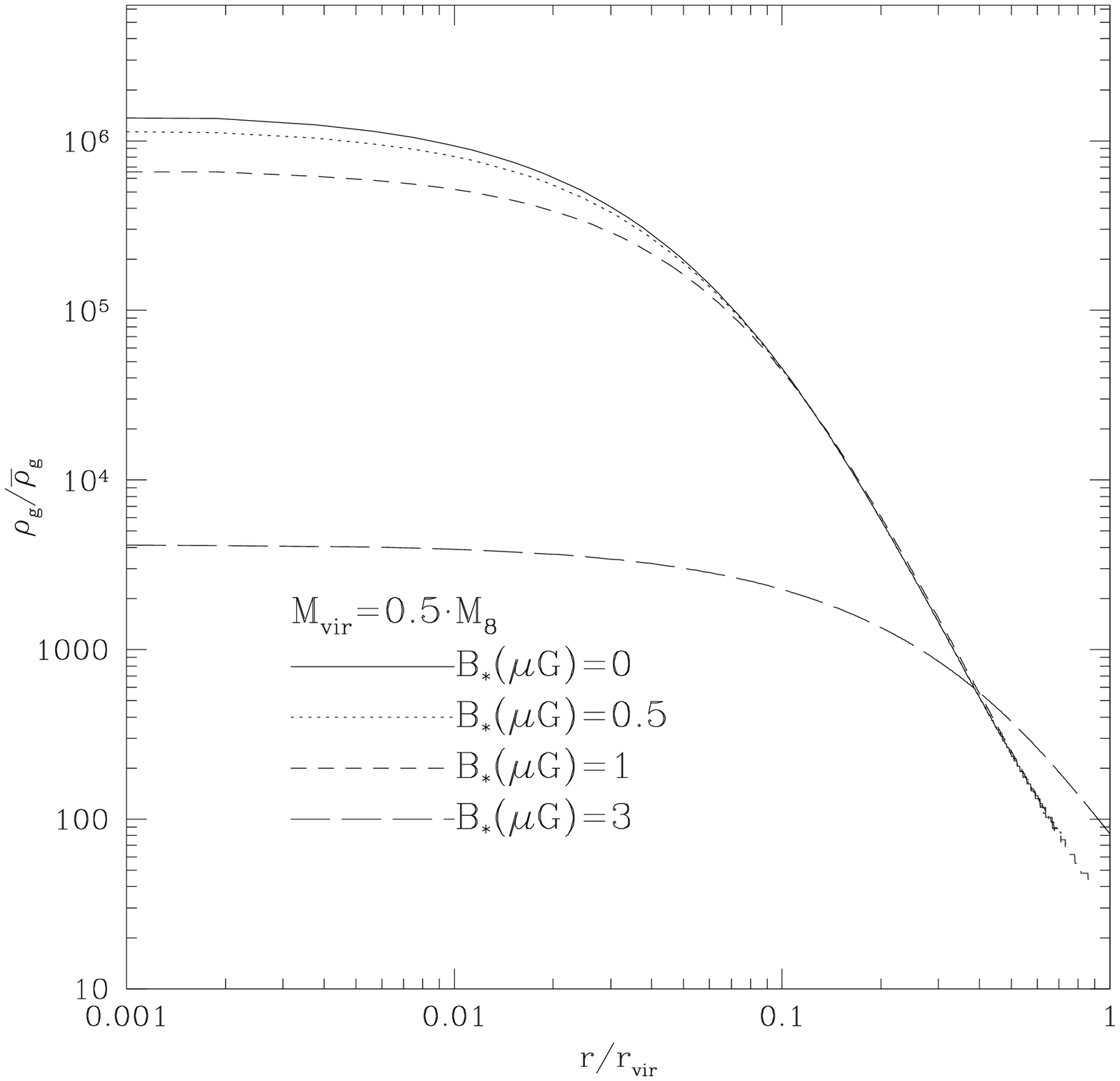,height=6.cm,width=6.cm,angle=0.0}
 \epsfig{file=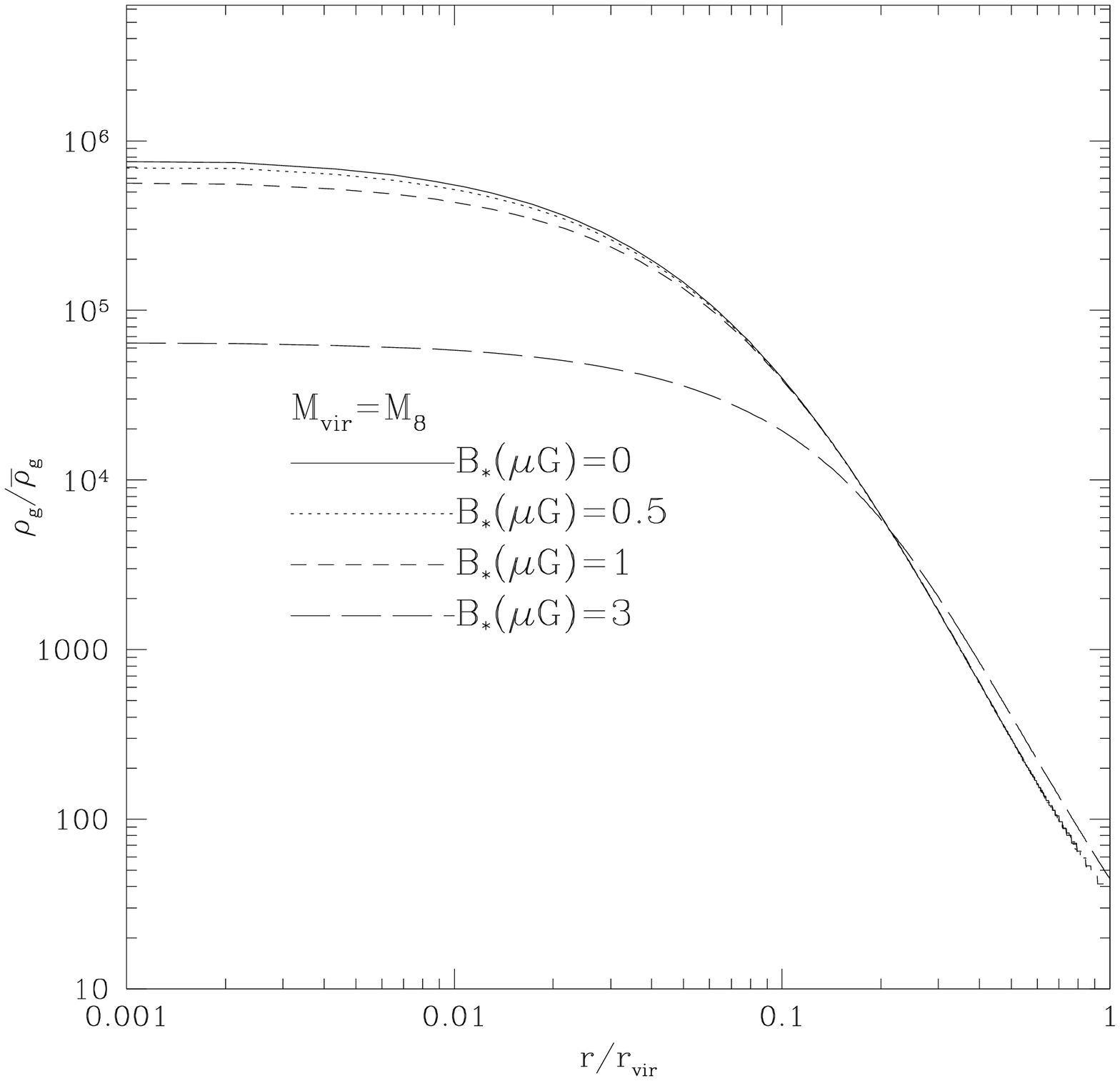,height=6.cm,width=6.cm,angle=0.0}
 \epsfig{file=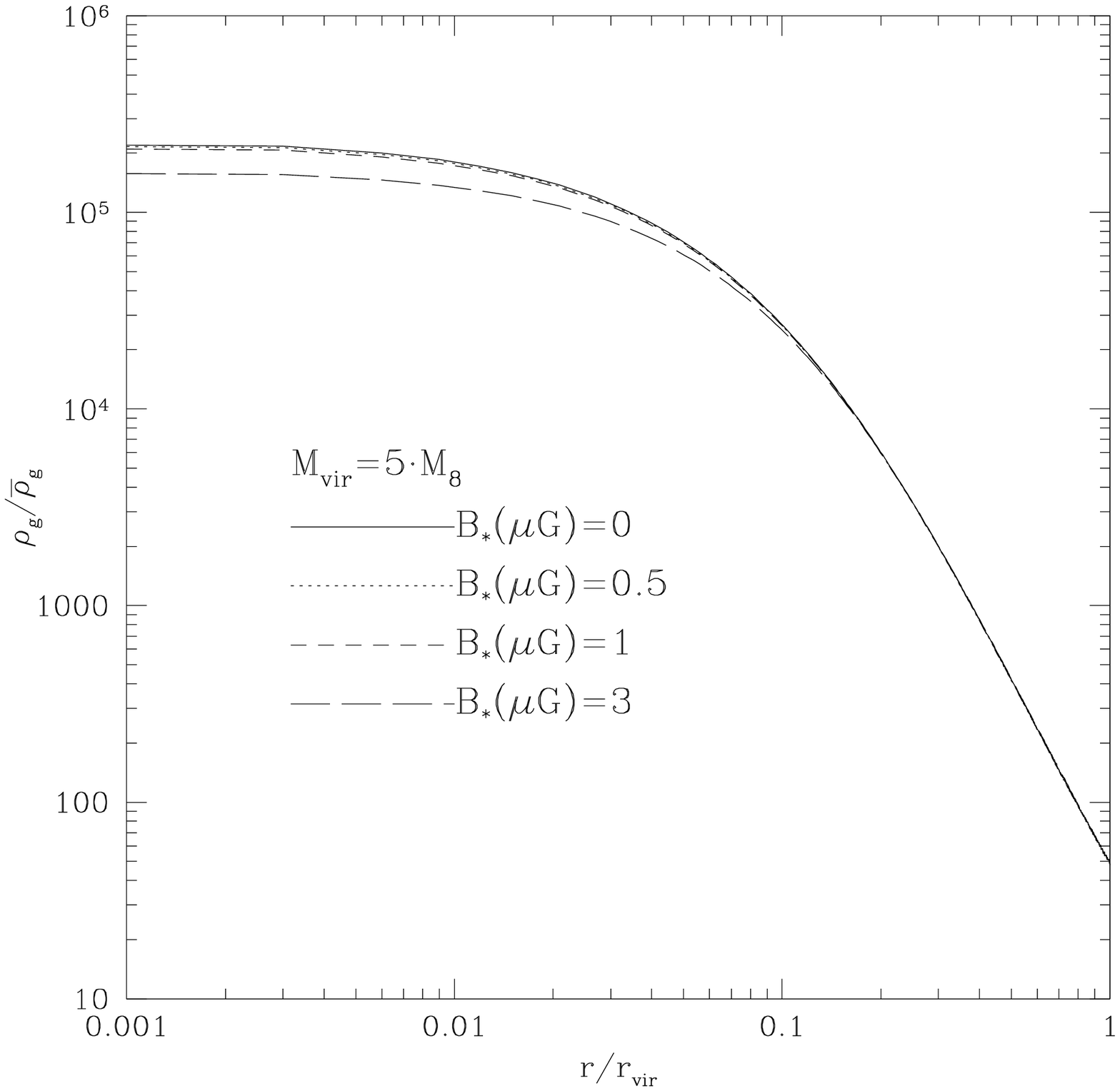,height=6.cm,width=6.cm,angle=0.0}
}
  \caption{\footnotesize{The IC gas density profile as derived from eq.(A19-A20)
  for three different cluster masses (as labelled in the three panels), and
  for values of the magnetic field $B_{*}=$ 0 (solid), 0.5 (dotted), 1 (short dashes)
  and 3 (long dashes) $\mu$G.
  The density profiles are normalized to the total gas mass at $r_{vir}$.
  }}
  \label{fig.rho_b_norm}
\end{center}
\end{figure}

\begin{figure}[!h]
\begin{center}
 \epsfig{file=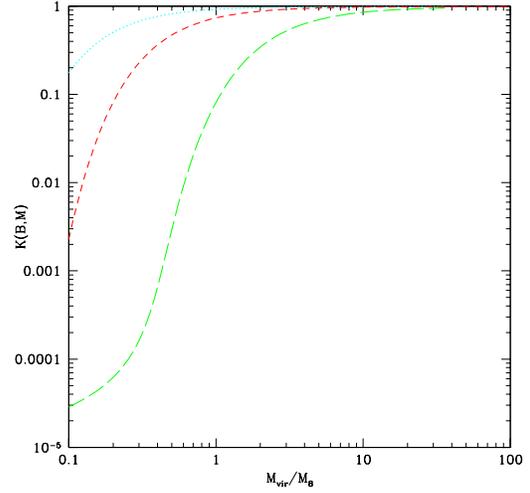,height=7.cm,width=7.cm,angle=0.0}
  \caption{\footnotesize{The function $K(M,B)= \rho_g(r,B)/ \rho_g(r,B=0)$,
  evaluated at $r=0.1 r_{200}$, is shown as a function of the cluster mass and for three
  different values of the magnetic field $B_{*}=$ 0.5 (cyan), 1 (red) and 3 (green) $\mu$G.
  }}
  \label{fig.kappa}
\end{center}
\end{figure}

Under these assumptions, we derived numerically the solution of the HE equation for the
IC gas density $\rho_g(r,B)$. Fig.\ref{fig.rho_b_norm} shows the density profile of the
IC gas derived from the HE condition and normalized to the cluster gas mass evaluated at
the virial radius as a function of the radius for different values of the magnetic field
$B_*$. The effect of the magnetic field is shown for various masses of the cluster
$M=0.5, 1$ and $5$ $M_8$, where $M_8$ is the mass enclosed in a $8 h^{-1}$ Mpc radius
spherical fluctuation at the epoch at which it goes non-linear: it takes the value $M_8
\approx 2 \cdot 10^{14} h^{-1}_{71} M_{\odot}$ in our reference cosmological model.
The B-field provides a pressure that tends to inflate the IC gas so that its final radial
profile extends outwards, in the case of density normalized to the central IC gas density
in the absence of the B-field.
Since observations and our modelling of the cluster structure provide a measure of the
gas distribution for a given mass (within a fixed radius), we study the effect of the
B-field for a fixed gas mass, i.e., extended, say, up to the virial radius, or more
generally to the radius $r_{\Delta}$ at which the density contrast is $\Delta$ (see the
Appendix for details).
Thus, the requirement to recover the fixed gas mass at the virial radius yields the
relation
 \be
 \rho_g(r,B) = K(M,B) \cdot \rho_g(r,B=0) \, ,
 \label{eq.rho_norm}
 \ee
with
 \be
 K(M,B) = {\int_0^c dx x^2 y_g(x,0) \over \int_0^c dx x^2 y_g(x,B)} \, .
 \label{eq.k}
 \ee
Such boundary conditions provide a decrease of the central gas density for increasing
values of $B_*$ (see Fig.\ref{fig.rho_b_norm}).
The effect of $B(r)$ clearly becomes milder for more massive systems due to the large DM
and gas mass contained there. In fact, the normalization function $K(M,B)= \rho_g(r,B)/
\rho_g(r,B=0)$ steeply rises from its low value at low masses and rapidly reaches unity
at mass scales that increase with increasing values of $B_*$ (see Fig.\ref{fig.kappa}).
Larger values of $B_*$, hence, increase the cluster mass range that is sensitive to the
effect of the magnetic field.

\begin{figure}[!h]
\begin{center}
\vbox{
 \epsfig{file=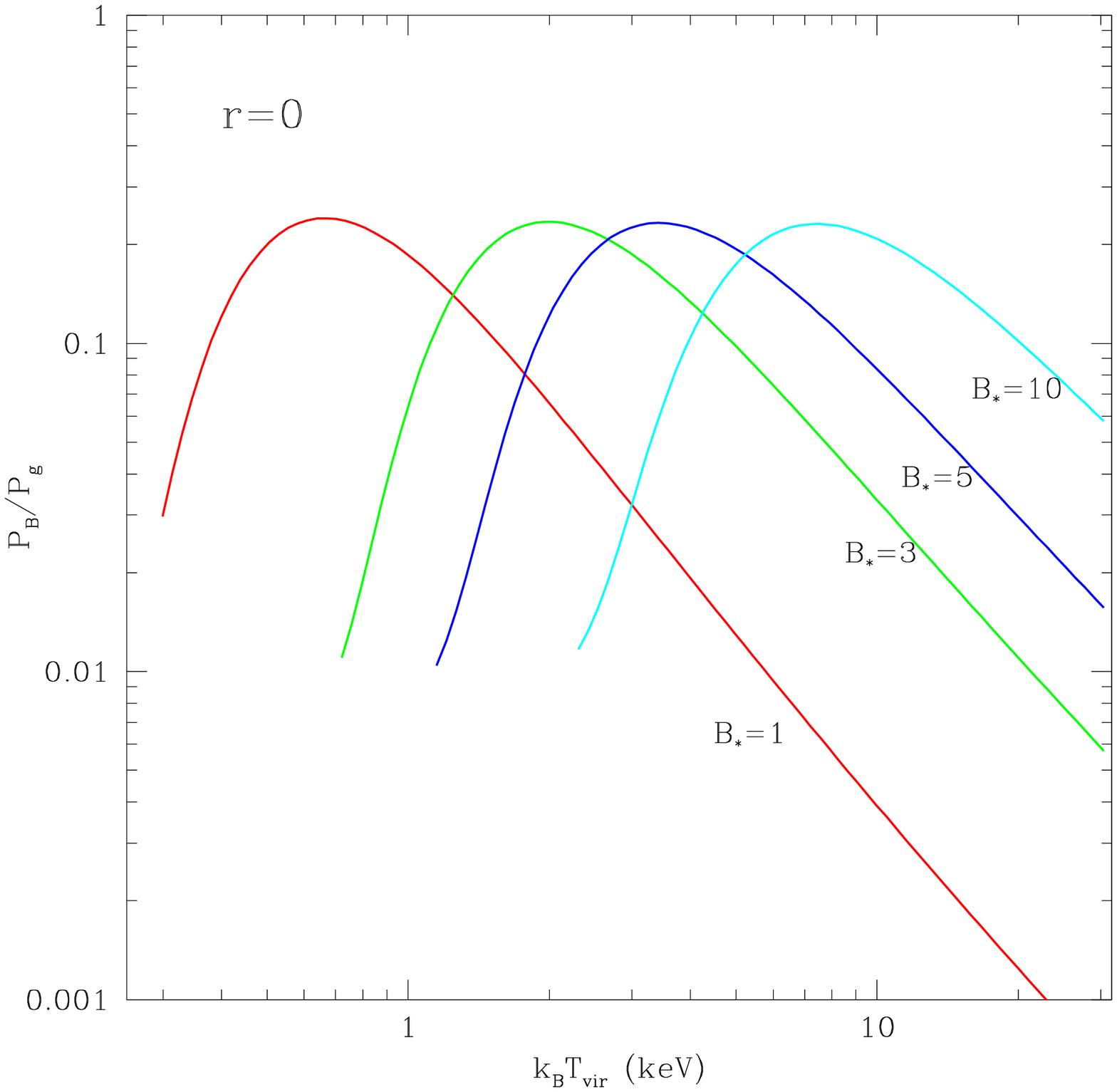,height=6.cm,width=6.cm,angle=0.0}
 \epsfig{file=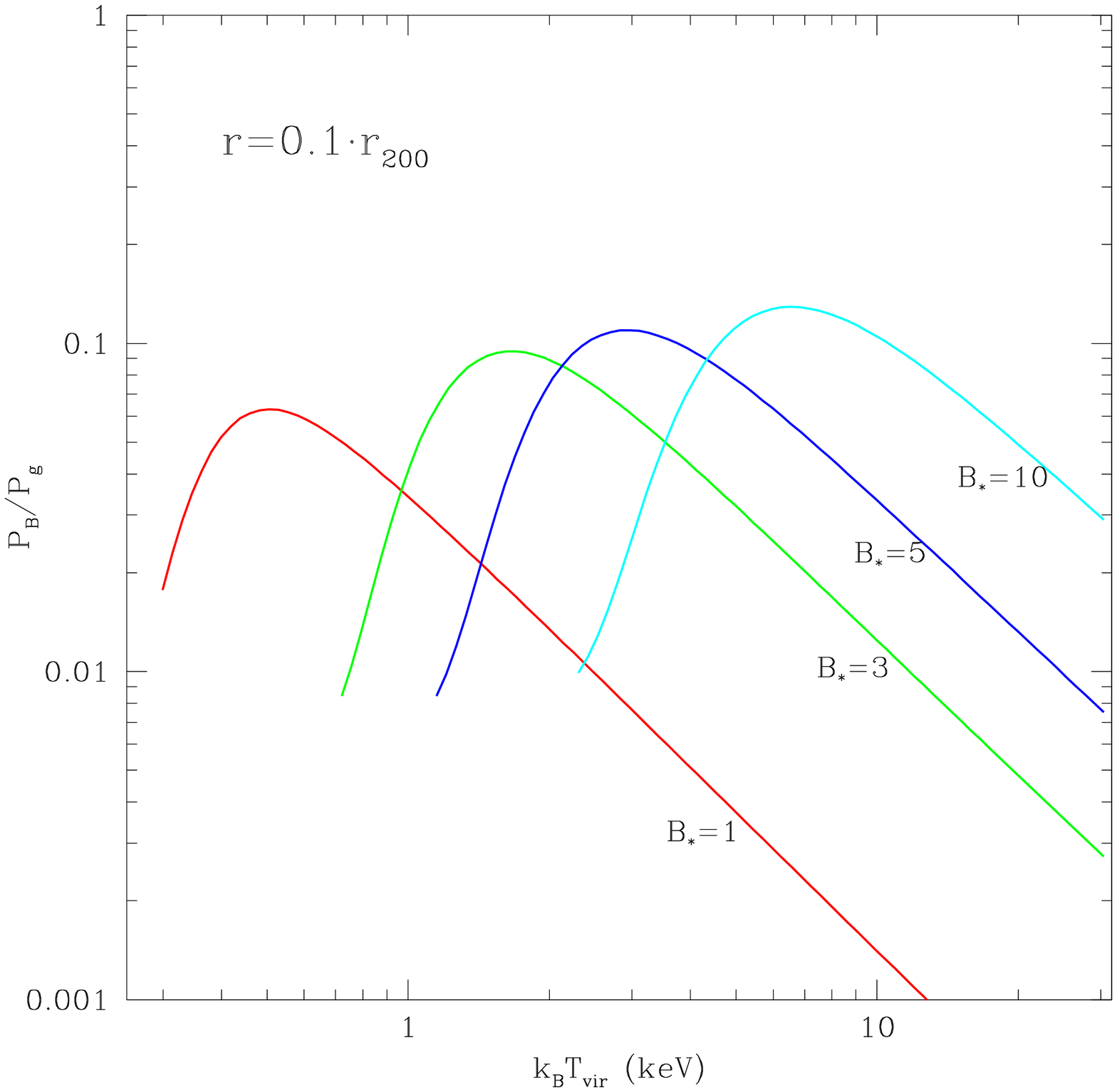,height=6.cm,width=6.cm,angle=0.0}
 \epsfig{file=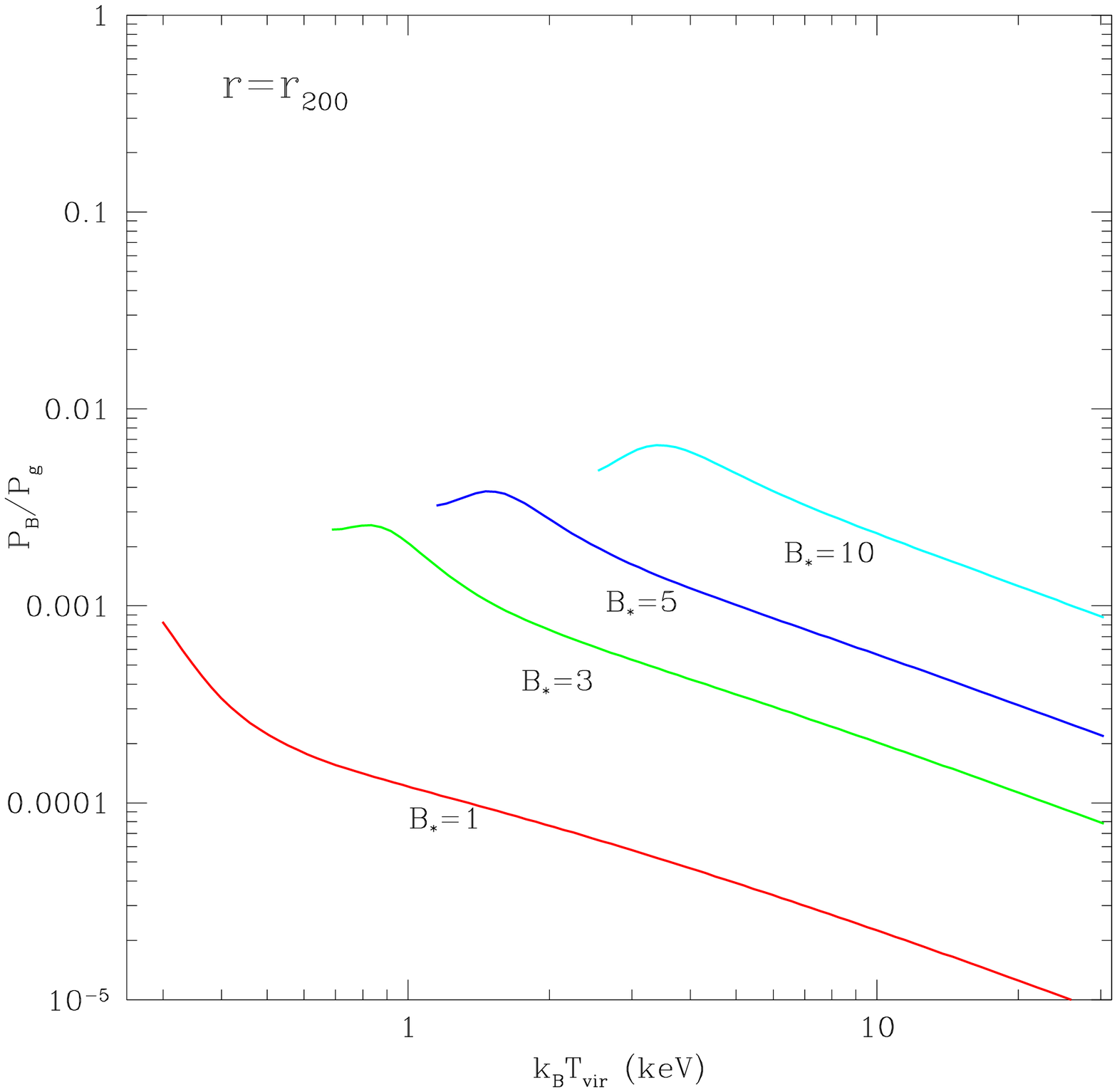,height=6.cm,width=6.cm,angle=0.0}
}
  \caption{\footnotesize{We show the ratio $P_B/P_g$ between the IC gas pressure
  and the magnetic pressure as a function of the cluster temperature for different values of the
  magnetic field $B_{*}=$ 1 (red), 3 (green), 5 (blue) and 10 (cyan) $\mu$G.
  The ratio $P_B/P_g$ is shown at three different cluster radii: $r=0$ (upper
  panel), $r=0.1 r_{200}$ (mid panel) and $r=r_{200}$ (lower panel).
  }}
  \label{fig.pressure}
\end{center}
\end{figure}

The ratio of the magnetic to IC gas pressure, $P_B/P_g$, corresponding to the density
profiles shown in Fig.\ref{fig.rho_b_norm} are calculated at three different radii $r=0$,
$0.1 r_{200}$ and $r_{200}$ and are shown in Fig.\ref{fig.pressure} as a function of the
cluster temperature.
The ratio between the magnetic pressure, $P_B \propto 8 \pi B^2(r)$, and the IC gas
pressure, $P_G \propto \rho_g T_g$, scales as $P_B/P_g \propto B_*^2 \rho_g^{2 \alpha -1}
/ T_g \sim B_*^2 K^{2 \alpha -1} / T_g$ according to our assumption for $B(r)$ (see
eq.\ref{magn0}).
For relatively high masses, when the function $K = \rho_g(r,B)/\rho_g(r,B=0) \to 1$ (see
Fig.\ref{fig.kappa}), the pressure ratio $P_B / P_g \propto B_*^2 T_g^{-1}$ and its
amplitude scales with the value of $B_*^2$.
For low masses the gas density is strongly affected by the magnetic field and hence the
function $K$ strongly decreases with decreasing $M$ so that the pressure ratio decreases
with decreasing $M$ (or $T$) as $P_B / P_g \propto K^{2 \alpha -1} T^{-1}$
At an intermediate mass (temperature) between the low-T tail (where $P_B/P_g$ strongly
decreases) and high-T tail (where $P_B/P_g$ asymptotically decreases as $1/T_g$), the
pressure ratio attains its maximum (following the curvature of the function $K$). The
temperature (mass) location of the maximum of $P_B/P_g$ depends of the value of $B_*$
reflecting the behaviour of the function $K$ as a function of $B_*$, i.e. the behaviour
of the gas density profiles shown in Fig.\ref{fig.rho_b_norm}.
In our specific model for the density and temperature of the IC gas in the presence of a
magnetic field, the value $P_B/P_g$ is $\sim 25 \%$ at most and peaks at increasing
values of $T$ for increasing amplitude of $B_*$.
The specific values of the maximum of $P_B/P_g$, of its temperature (mass) location and
of the low-T tail depend on the specific choice for the exponent $\alpha$ in
eq.(\ref{magn0}) and are not universal. Nonetheless, the general trend of the pressure
ratio shown in Fig.\ref{fig.pressure} is preserved for the specific form of the radial
dependence of the magnetic field as in eq.(\ref{magn0}) and for values of $\alpha > 0.5$,
as indicated by the available data.

\subsection{The magnetic virial theorem for galaxy clusters}
 \label{sect.mvt}

The temperature of the IC gas that is used in the HE condition is consistently derived,
in the framework of our isothermal model, from the magnetic virial theorem (MVT). We
recall here the main aspects of the IC gas temperature derivation from the  MVT following
Colafrancesco \& Giordano (2006a).\\
Under the assumption of an ICM in hydrostatic equilibrium with the potential well of a
spherically-symmetric, isolated, virialized and magnetized cluster, the general relation
between the ICM temperature $T_g$ and the cluster virial mass $M_{vir}$ is obtained by
applying the MVT that, for a static and isothermal galaxy cluster, reads
 \be
 2K + 2U + U_B + W = 0 \, .
 \label{eq.mvt}
 \ee
Here $U_B$ is the magnetic energy of the system, $U$ is the kinetic energy of the gas (we
assume that the Dark Matter is cold and collisionless), $K$ is the Dark-Matter particle
kinetic energy and $W$ is the total potential energy (see Colafrancesco \& Giordano 2006a
for details). The previous eq. (\ref{eq.mvt}) holds specifically in the absence of an
external medium. For the more general case of a cluster that is immersed in a Inter
Galactic Medium (IGM) or in an external medium that exerts an external pressure
$P_{ext}$, eq.(\ref{eq.mvt}) yields the formula for the temperature of the gas in virial
equilibrium
 \be
 \frac{k_BT_g}{\mu m_p}=\frac{\xi
 G}{3}\frac{M_{vir}}{r_{vir}}\left(1-\frac{M_{\phi}^2}{M_{vir}^2}+
 \frac{P_{ext}}{P_{vir}}\right),
 \label{eq.TvirM}
 \ee
where $P_{vir} = (\frac{4\pi}{\xi G}\frac{r_{vir}^4}{M_{vir}^2})^{-1}$ with usually $\xi
\simgt 1$, and $M_{\phi}$ is given by
 \be
 M_{\phi}\simeq 1.32\cdot 10^{13} M_{\odot} \left[\frac{I(c)}{c^3} \right]^{1/2}
 \left(\frac{B_*}{\mu G}\right)\left(\frac{r_{vir}}{Mpc}\right)^2 \, .
 \label{eq.mb}
 \ee
Here $I(c)=\int_0^c (\rho_g(r=0)/{\bar \rho_g}(z=0))^{2\alpha} x^2 y_g^{2\alpha}(x,B=0)
dx$, and
$y_g(x,B=0) = \rho_g(x)/\rho_g(x=0)$ is the gas density profile normalized to the central
gas density (i.e. the solution of the hydrostatic equilibrium equation in the absence of
a magnetic field, see Appendix for details).\\
For the case $P_{ext}=0$ and $B=0$, one finds $M_{\phi}=0$ and the well-known relation
 \be
 k_B T_g(B=0)=-{ \xi \mu m_p W \over 3 M_{vir}}
 \label{eq.t_b0}
 \ee
re-obtains (here $\mu=0.63$ is the mean molecular weight, corresponding to a hydrogen
mass fraction of $0.69$, $m_p$ is the proton mass and $k_B$ is the Boltzmann constant).\\
For $B > 0$, one finds $M_{\phi} > 0$ and the temperature of the gas at fixed $M_{vir}$
reads
 \be
 k T_g = k T_g(B=0) \left(1-\frac{M_{\phi}^2}{M_{vir}^2}+
 \frac{P_{ext}}{P_{vir}} \right) \, .
 \label{eq.TvirM_norm}
 \ee
The temperature of a cluster calculated in the presence of a magnetic field (with
$P_{ext}=0$) is lower than that given by eq.(\ref{eq.t_b0}) because the additional
magnetic field energy term $U_B$ adds to the MVT.\\
The presence of an external pressure $P_{ext}$ tends to compensate the decrease of $T_g$
induced by the magnetic field.
In the following, we consider  values of $P_{ext}$ (which are relevant for the structures
we consider here) as estimated in the IGM and in the rich cluster outskirts.
For values of the temperature, $T_{IGM} \sim  10^6 $K, and density, $n_{IGM} \sim 10^{-5}
cm^{-3}$, as estimated by the WHIM structure around large-scale overdensities (see, e.g.,
Fang \& Bryan 2001), $P_{ext} \sim 1.7\cdot10^{-3}$ eV cm$^{-3} (n_{IGM}/10^{-5}
cm^{-3})(T_{IGM}/2 \cdot 10^6 K)$.
However, in the outer regions of massive clusters (at $r \simgt r_{vir}$, where the main
source of pressure could be due to the momentum of the infalling gas accreting onto the
cluster), studies of the mean projected temperature profile for the cluster sample
derived by Piffaretti et al. (2005, see their Fig.4) indicate that the external gas
pressure can reach values up to $P_{ext} \sim 0.2$ eV cm$^{-3} (n/10^{-4}
cm^{-3})(T_g/1.7 \cdot 10^7 K)$ (we consider here a typical rich cluster with $T_g = 10$
keV). In the case of such a hot cluster, the value of $P_{ext}$ is a significant
fraction, $\sim 4 \%$, of the central ICM pressure and $\sim 50 \%$ of the ICM pressure
at the virial radius for a typical cluster.
However, for lower-temperature clusters, such values of $P_{ext}$ are expected to be
sensitively lower.
A value $P_{ext} \sim 0.2$ eV cm$^{-3}$, as estimated at the outskirts ($r \simgt
r_{vir}$) of rich, hot clusters, can be considered, hence, as an upper bound to
$P_{ext}$; in addition, an exact determination of the total cluster mass (which is
subject to various systematic uncertainties, see, e.g., Rasia et al. 2006) requires an
extension beyond $r_{vir}$.
Lower values of $P_{ext}$, down to the value found in the WHIM, have progressively less
importance for the MVT.\\
For reasonable values of $B_* \simgt$ a few $\mu$G, the quantity $M^2_{\phi}
> M_{vir}^2 \cdot (P_{ext}/P_{vir})$ in eq.(\ref{eq.TvirM}) and the main effect is a
reduction of the cluster temperature which is more pronounced for less massive systems,
where $ M_{\phi}$ becomes comparable to $M_{vir}$. The effects of the magnetic field and
of the external pressure are greater for low-$M$ clusters (see Colafrancesco \& Giordano
2006a and Fig.\ref{fig.MT_B}).


\section{The radial entropy profiles of magnetized clusters} \label{sect.sprofile}

\begin{figure}[!h]
\begin{center}
\vbox{
 \epsfig{file=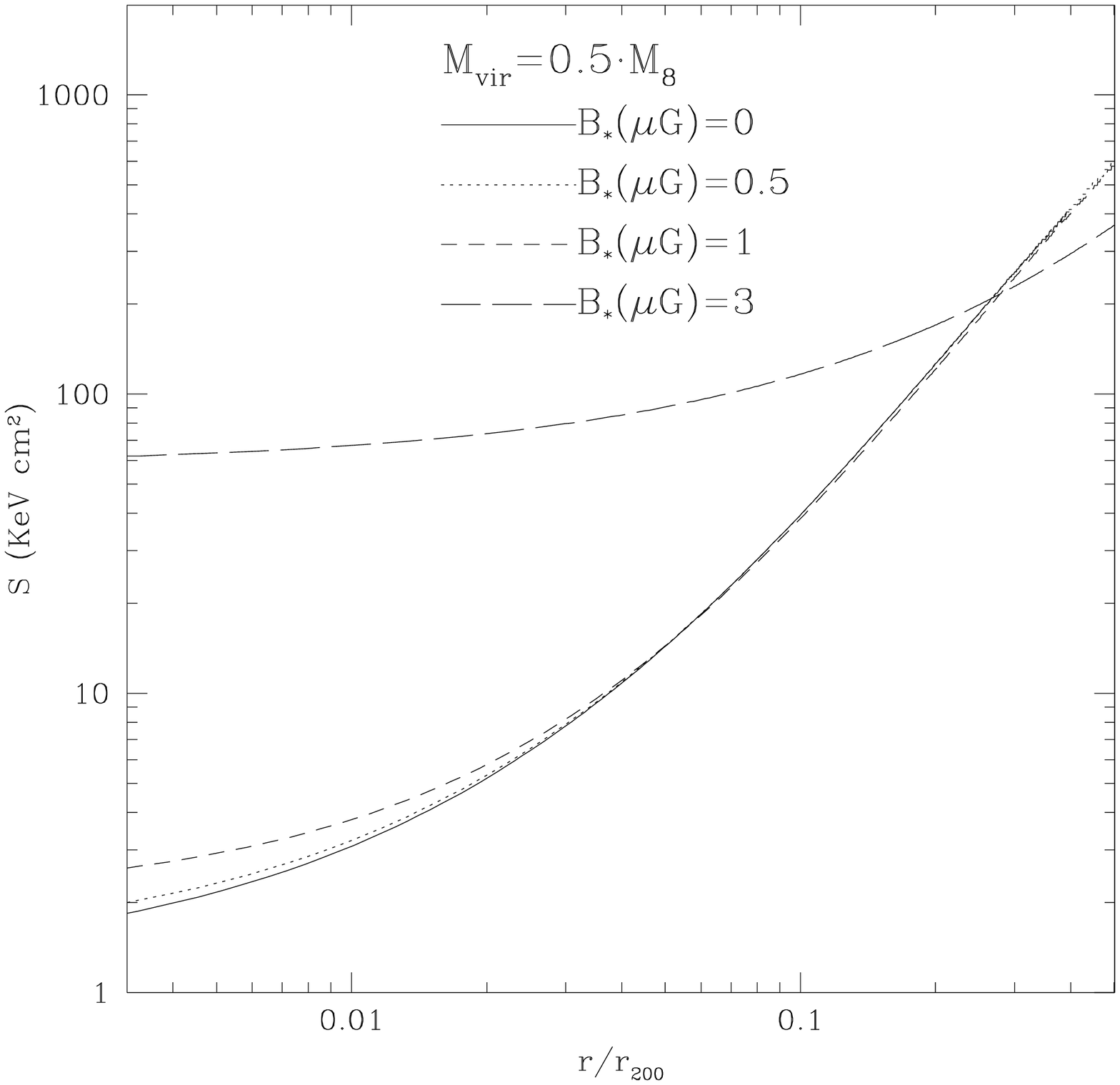,height=6.cm,width=6.cm,angle=0.0}
 \epsfig{file=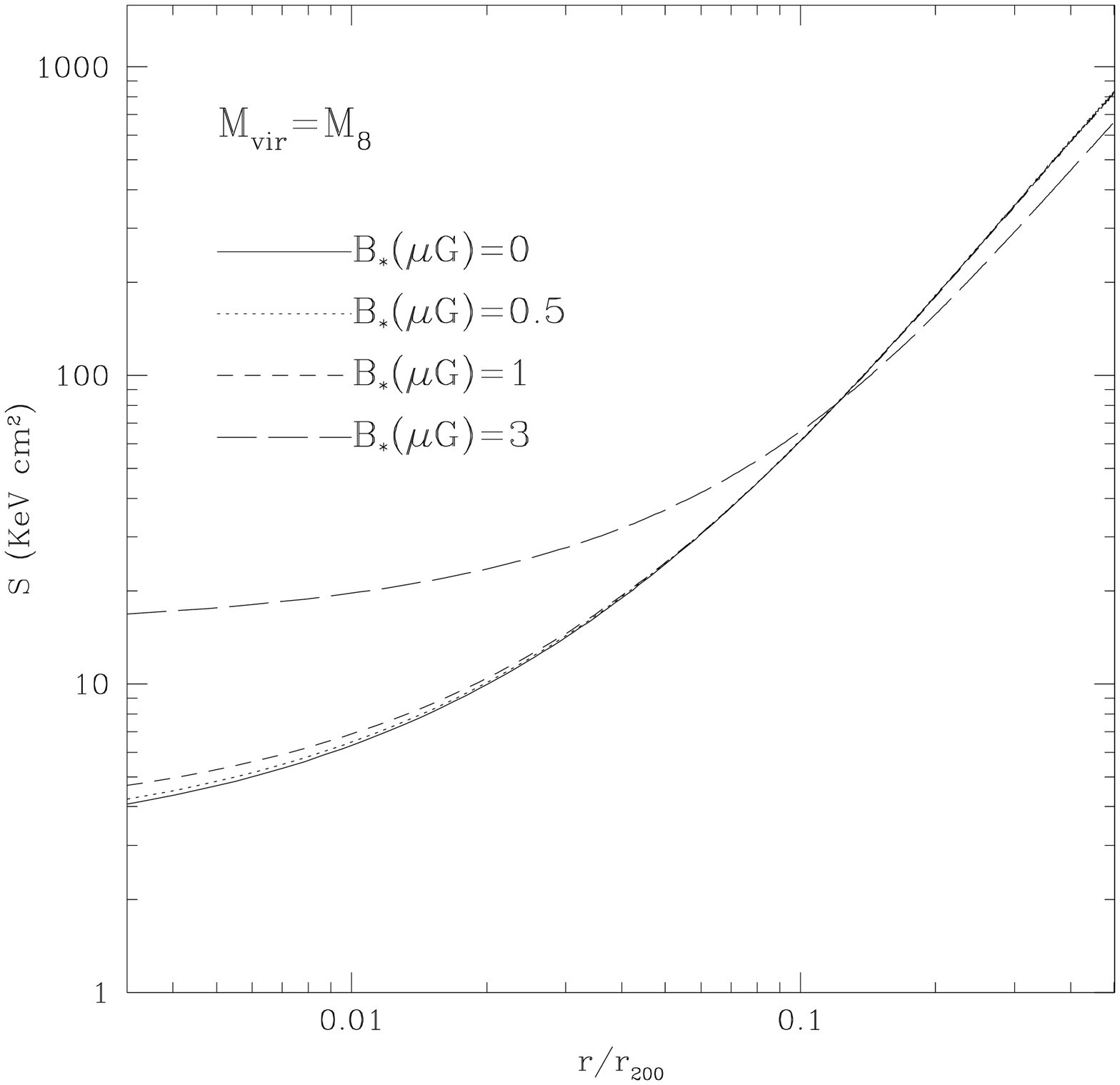,height=6.cm,width=6.cm,angle=0.0}
 \epsfig{file=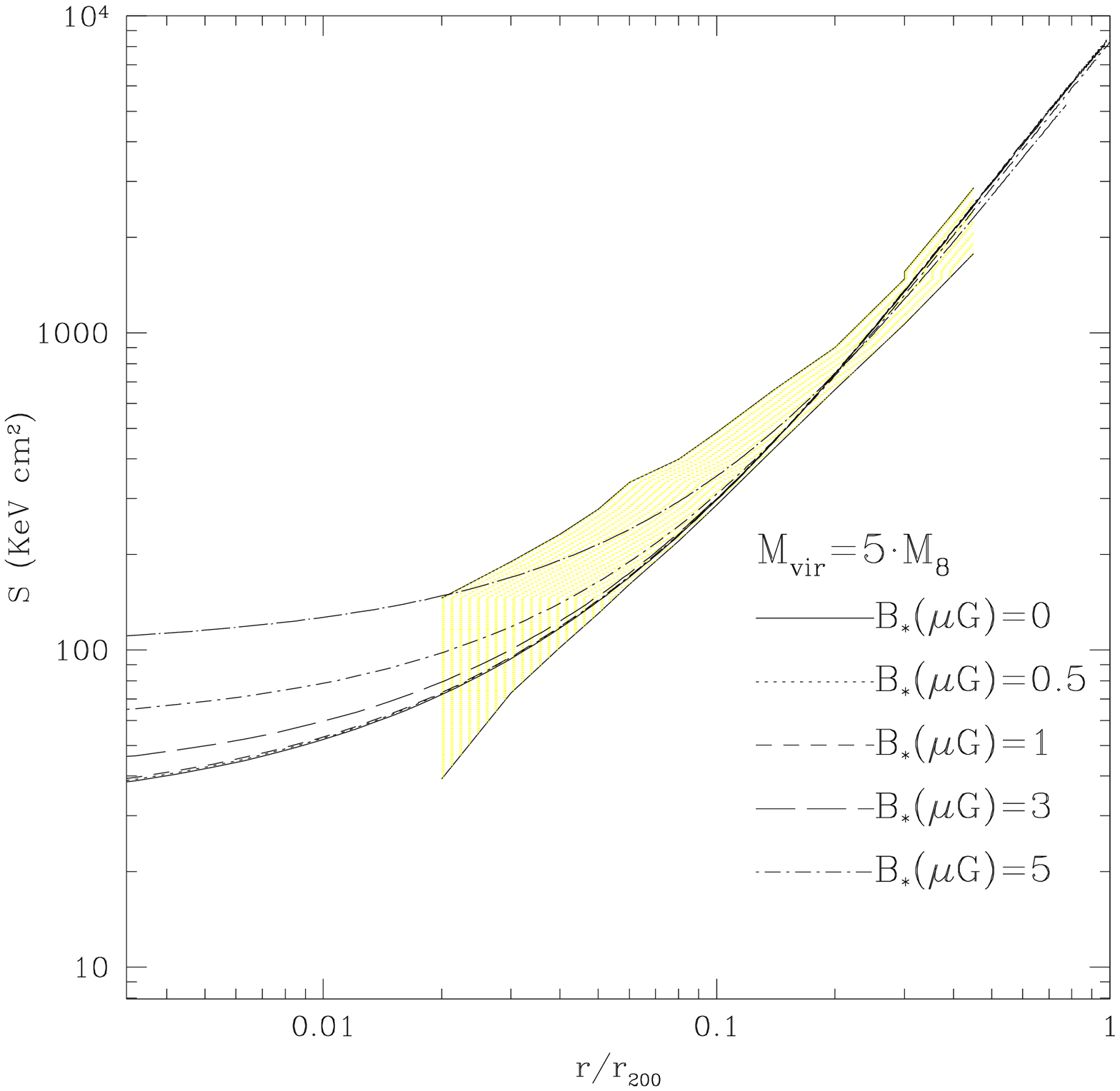,height=6.cm,width=6.cm,angle=0.0} }
  \caption{\footnotesize{IC gas entropy profile for various values of the
  magnetic field $B_{*}= 0, 0.5, 1$ and $3$ $\mu$G and for three different cluster masses:
  $0.5 M_8$ (top panel), $M_8$ (mid panel) and $5 M_8$ (bottom panel).
  The density profiles are normalized to the total gas mass at $r_{vir}$.
  The shaded region in the last plot for a cluster with $M_{vir} = 5 M_8$ show the
  1 sigma area spanned by the normalized entropy profiles derived by
  Pratt et al. (2006) for 10 nearby clusters.
  }}
  \label{fig.s_b_norm}
\end{center}
\end{figure}

Given the previous results on the temperature and on the gas density in the presence of a
B-field, we can now discuss the impact of the magnetic field on the entropy profile of
galaxy clusters.\\
The entropy of the IC gas is usually defined as
 \be
 S = {k_B T_g \over \rho_g^{2/3}}
 \label{eq.entropy}
 \ee
and, in our model, it depends on the magnetic field through the B-dependence of the
cluster temperature $T_g(B)$, as derived from the MVT (see eq.\ref{eq.mvt}), and from the
IC gas density $\rho_g(r,B)$, as derived from the HE condition (see eqs.A19-A20).\\
Our model provides a simple but consistent description of the B-dependence of the
(isothermal) cluster entropy and contains all the relevant physical information. However,
the detailed effect of the magnetic field on the cluster entropy should be addressed in a
truly magneto-hydro-dynamical (MHD) model of cluster formation and evolution, which is
able to catch the details of the turbulent amplification and of the evolution of the
B-field associated with the shocked gas in the cluster environment. Such a task is well
beyond the scope of this paper and could be better tackled through MHD cosmological
simulations and/or semi-analytical methods that will be addressed elsewhere. Nonetheless,
the model we present here is able to describe the overall properties of cluster entropy,
namely its radial behaviour and its correlation with the cluster temperature.
In Fig.\ref{fig.s_b_norm} we show the radial entropy profiles of the IC gas calculated by
using the IC gas density profiles normalized to the cluster mass gas (these are shown in
Fig.\ref{fig.rho_b_norm}).
The entropy profiles we obtain flatten in the inner cluster region for increasing values
of the magnetic field $B_*$. This behaviour reflects the decrease of the central IC gas
density for increasing values of the magnetic field, as shown in
Fig.\ref{fig.rho_b_norm}.
The bottom panel in Fig.\ref{fig.s_b_norm} shows also the comparison of our entropy
profiles for a typical $M_{vir}=5 M_8$ cluster with the region enclosed by the mean
plus/minus the 1 $\sigma$ standard deviation of the scaled entropy profiles for ten
clusters observed with XMM (Pratt et al. 2006).
These clusters have, in fact, quite high masses around the reference value we consider
here. The predictions of our model agree quite well with the data, and the widening of
the observed entropy profiles at small radii could reflect the dispersion of the
B-dependent density profiles found for high values of $B_*$. In this respect, the XMM
cluster data, taken at face value, limit the amplitude of the magnetic field to values
$B_* \simlt 7$ $\mu$G for this mass range.\\
The mean temperature used to investigate the scaling properties of the XMM clusters has
been estimated in the region $0.1 r_{200} < r < 0.5 r_{200}$ and it is quite constant for
most of the XMM clusters (see Pointecouteau et al. 2005). This fact allows us to directly
compare the predictions of our isothermal model with the observed entropy profiles
derived by Pratt et al. (2006).\\
Furthermore, our model predicts a sensitive flattening of $S(r)$ at smaller radii $< 0.2
r_{200}$ which is more evident for higher values of $B_*$. This behaviour is similar to
the entropy flattening shown in the inner cool cores of the clusters studied with Chandra
(see, e.g., Donahue et al. 2006). Note, however, that a closer comparison of the
(isothermal) model we consider here with these cool cluster cores (where one observes
temperature jumps of the order of a factor of 2-3 with respect to the outer temperature,
see, e.g., Colafrancesco, Dar \& DeRujula 2004 and references therein) cannot be done
straightforwardly, and it demands a more detailed modeling of the magnetized core regions
that will be presented elsewhere (Colafrancesco \& Giordano 2006b).


\section{The M-T relation for magnetized clusters}
 \label{sec.mt}

\begin{figure}[h!]
\begin{center}
 \epsfig{file=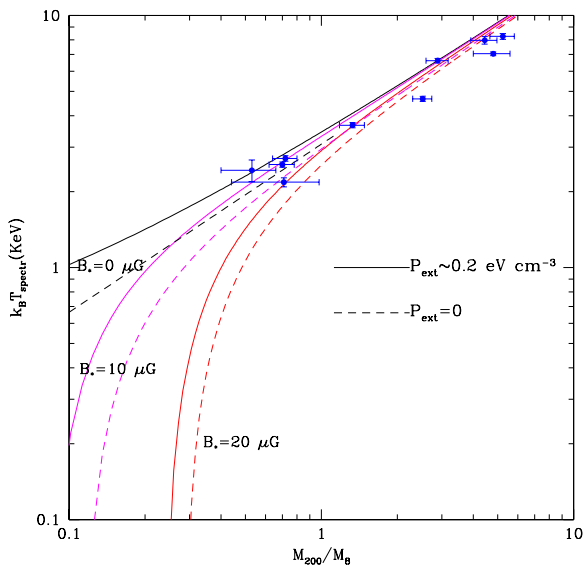,height=8.cm,width=8.cm,angle=0.0}
  \caption{\footnotesize{The $T_{spectr}-M_{200}$ relation at $z=0$ for clusters
  that contain a magnetic field $B_*=0$ (black), $10$ (magenta) and $30$ $\mu$G (red),
  and for values of $P_{ext}=0$ (dashed curves) and $P_{ext}=0.2$ eV cm$^{-3}$ (solid curves).
  Data are taken from Arnaud (2005).
  }}
  \label{fig.MT_B}
\end{center}
\end{figure}

To derive consistent structural relationships for clusters and groups, we first need to
derive their $M-T$ relation.
The $M-T$ relation is a basic structural relation for groups and clusters and it is the
starting point to study the entropy and X-ray luminosity scaling with $T$ and/or $M$.
The $T-M$ relation for magnetized clusters has been derived by Colafrancesco \& Giordano
(2006a) and is also shown in Fig.\ref{fig.MT_B} where it is compared to the data for the
XMM clusters used to derive the entropy profiles previously discussed (see Pratt et al.
2003).
Our isothermal model is consistent with the previous data because these are obtained by
using an overall spectroscopic temperature (which is constant in the radial range
$(0.1-0.5) r_{200}$ and also because the values of $T_{spectr.}$ have been obtained by
performing an isothermal fit of the cluster spectra obtained in the same $(0.1-0.5)
r_{200}$ radial range.\\
The normalization of the observed $T_{spectr.}-M_{200}$ relation is however found to be
discrepant by $\sim 30 \%$ with respect to the value derived from numerical simulations
including only gravitational heating (see, e.g. Evrard et al. 1996), a well known problem
(see, e.g., Arnaud 2005 for a review) which usually requires one to re-normalize the
predictions of analytical models.
In our model we normalize the $T_{spectr.}-M_{200}$ relation for the case $B=0$ to the
observed data derived by Arnaud et al. (2005) by assuming $T_{spectr} = T_g$, where $T_g$
is obtained from eq.(\ref{eq.TvirM}), and a temperature boost of a factor $\xi \approx
1.5$ in eq.(\ref{eq.TvirM}), as can be expected from the continuous shock-heating of the
IC gas within the virial radius after the formation of the original structure (see, e.g.,
Makino et al. 1998, Fujita et al. 2003, Ryu et al. 2003).
Such a prescription for $\xi$ has been used by the previous authors for unmagnetized
clusters in the absence of external pressure.
A systematic effect which tends to increase the cluster temperatures is the presence of a
minimal external pressure in eqs.(9) and (12). A value $P_{ext} \sim(0.1-0.2) P_{vir}$
(like that found in the IGM around clusters) could easily accommodate for an overall
boost value of $ \xi \approx 1.5 \times  (P_{ext} / P_{vir}) > 1.5$ and reasonably in the
range $1.65 - 1.8$, for the previous values of $P_{ext}$.
Given the large theoretical uncertainty on the non-gravitational heating efficiency, we
adopted an overall value $\xi \approx 1.8$ to normalize our prediction to the data point
at $M_{200}  \approx 3 \cdot M_8$ in Fig.5, which is the point with the smaller intrinsic
error. However, values of $\xi$ in the plausible range $1.5 - 1.8$ marginally change our
predictions.\\
The relation $M_{200} \simeq 0.77 M_{vir}$ is also found in our mass scale definition.\\
Small variations of temperatures with respect to their unmagnetized values are found for
massive clusters since $M_{\phi} \ll M_{vir}$ in this mass range and the value of
$P_{ext}$ has little or negligible effect (see Fig.\ref{fig.MT_B}).
This is in agreement with the results of numerical simulations (Dolag et al. 2001a).
However, when $M_{\phi}$ becomes comparable to $M_{vir}$, the IC gas temperature becomes
lower than its unmagnetized value and the $T-M$ relation steepens in the range of less
massive systems like groups and poor clusters. The temperature $T_g$ formally tends to
zero when $M_{\phi} \to M_{vir}(1+P_{ext}/P_{vir})^{1/2}$. However, this limit is
unphysical since it corresponds to an unstable system in which the magnetic pressure
overcomes the gravitational pull. Thus, any physical configuration of magnetized
virialized structures must have $M_{\phi} < M_{vir}(1+P_{ext}/P_{vir})^{1/2}$.\\
The effect of $P_{ext}$ counterbalances the effect of the $B$-field on the $T-M$
relation, and its amplitude increases for low-$M$ systems. (see Fig.\ref{fig.MT_B}; see
also Colafrancesco \& Giordano 2006a for details).


\section{The $S-T$ relation for magnetized clusters}
\label{sec.st}

Using the $T-M$ relation presented in Sect.\ref{sec.mt}, we can now derive our
predictions for the $S-T$ relation for clusters with a magnetic field.
Here, we will compare our models for the IC gas distribution in the presence of a
magnetic field with the data of the $S-T$ relation (see Ponman et al. 2003, see also
Arnaud 2005 for a review). The entropy in this correlation has been evaluated at a
distance $0.1 r_{200}$ from the cluster center and thus we have to evaluate both the IC
gas density and its temperature at this radius.

\begin{figure}[!h]
\begin{center}
 \epsfig{file=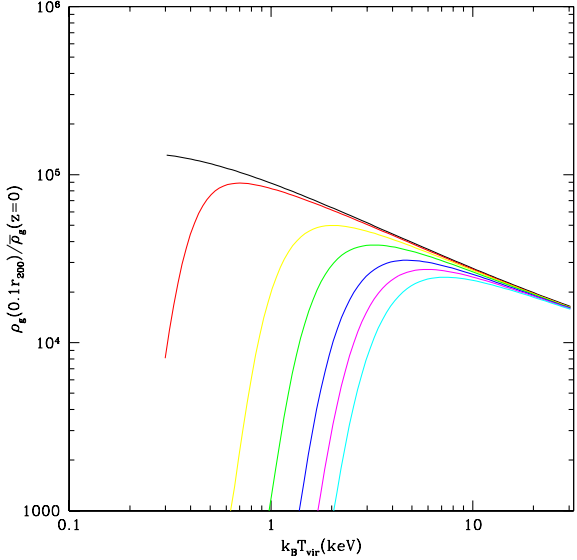,height=8.cm,width=8.cm,angle=0.0}
  \caption{\footnotesize{The temperature dependence of the IC gas density evaluated at
  $0.1 r_{200}$ for various values of the magnetic field $B_*=0$ (black curve), $0.5$ $\mu$G
  (red), $1$ $\mu$G (yellow), $2$ $\mu$G (green), $3$ $\mu$G (blue), $5$ $\mu$G (magenta) and
  $10$ $\mu$G (cyan).}}
  \label{fig.rho_0.1r200}
\end{center}
\end{figure}

\begin{figure}[!h]
\begin{center}
 \epsfig{file=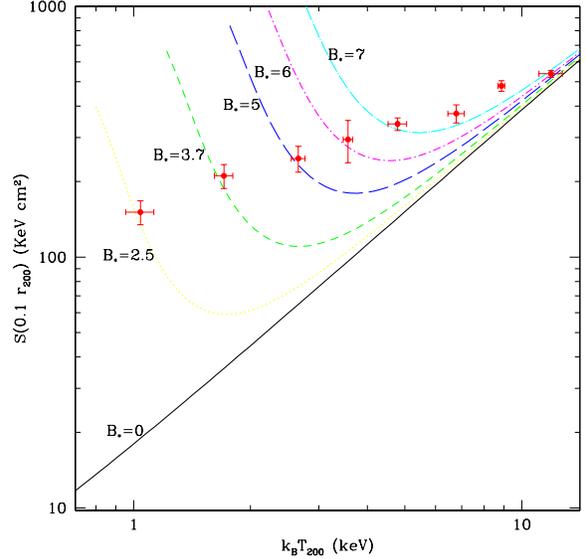,height=8.cm,width=8.cm,angle=0.0}
  \caption{\footnotesize{The entropy $S(0.1 r_{200})$ as a function of the temperature
  for various values of the magnetic field $B_*$ as labelled. The binned entropy data are taken
  from Ponman et al. (2003).
  }}
  \label{fig.st}
\end{center}
\end{figure}

\begin{figure}[!h]
\begin{center}
 \epsfig{file=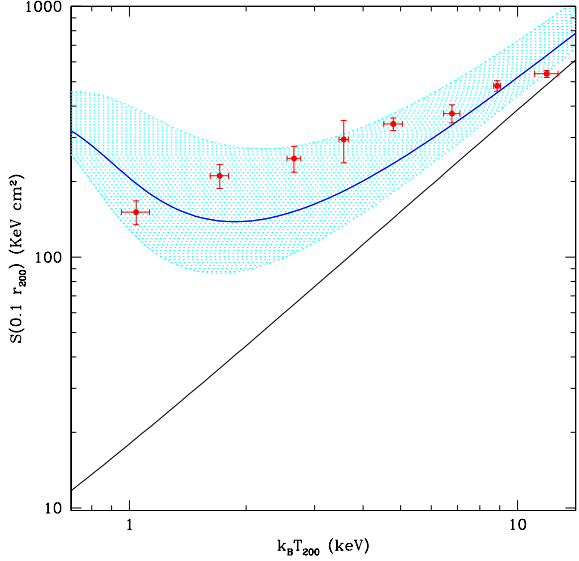,height=8.cm,width=8.cm,angle=0.0}
  \caption{\footnotesize{The entropy $S(0.1 r_{200})$ is shown as a function of the temperature
  for a magnetic field scaling $B_* \sim T_{200}^{\eta}$ with $\eta= 0.5$ (solid curve). The
  dashed area enclosed by the two entropy curves calculated for $B_* \sim T_{200}^{0.4}$
  and for $B_* \sim T_{200}^{0.6}$ shows the range of variation for $B_*$ allowed by the
  binned entropy data.
  }}
  \label{fig.st_fit}
\end{center}
\end{figure}

\begin{figure}[!h]
\begin{center}
 \epsfig{file=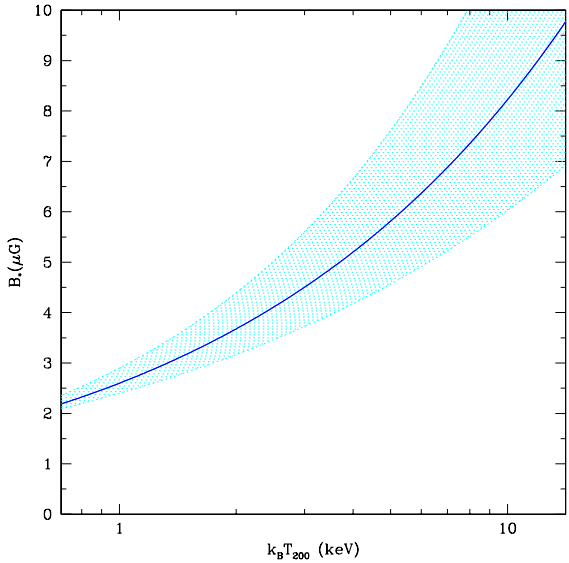,height=8.cm,width=8.cm,angle=0.0}
  \caption{\footnotesize{The function $B_* = 2.6 \mu G (T_{200}/keV)^{0.5}$ which best fits the
  $S-T$ relation in Fig.\ref{fig.st_fit} is shown by the solid curve. The shaded area shows the range of
  variation of $B_*(T_{200})$ enclosed by the curves $B_* = 2.4 \mu G (T_{200}/keV)^{0.4}$ (lower
  bound) and $B_* = 2.9 \mu G (T_{200}/keV)^{0.6}$ (upper bound).
  }}
  \label{fig.bt_fit}
\end{center}
\end{figure}

Due to the decrease of the IC gas density for increasing values of the magnetic field
(see Fig.\ref{fig.rho_b_norm}), the density of the IC gas evaluated at a distance $0.1
r_{200}$ exhibits a peculiar trend with increasing cluster temperature (mass) as shown in
Fig.\ref{fig.rho_0.1r200}.
The presence of a temperature dependence of the gas density $\rho_g(0.1 r_{200})$, even
in the case $B=0$, is a consequence of the HE condition (see eq.4), independently of the
chosen equation of state for the gas in the MVT. The specific solution of the HE
condition for the unmagnetized case ($B=0$) is given in eq.(A.10) of the Appendix. It is
clear that $\rho_g(r) \propto exp[-3 c/m(c) \int_0^x du m(u)/u^2]$ exhibits a temperature
dependence (like that shown in Fig.\ref{fig.rho_0.1r200}) since the concentration
parameter $c$ depends on $T_g$ through eq.(3).\\
For each fixed value of $B_*$, the presence of a B-field produces a sharp rise of the gas
density $\rho_g(0.1 r_{200})$ in the low-T region followed by a smooth decline at higher
temperatures where the density tends to the unmagnetized value (solid black curve). This
effect is produced by the physical boundary condition for the IC gas density, i.e. by the
requirement to normalize the IC gas density to the total gas mass at $r_{vir}$ for each
cluster.
The sharp rise of $\rho_g(0.1 r_{200})$ with the cluster temperature (mass) is produced
by the effect of the function $K(M,B)$ (see Fig.\ref{fig.kappa}): when $K \to 1$ at high
masses, then $\rho_g(0.1 r_{200},B)$ tends to its value for unmagnetized clusters with
$B=0$ (i.e., the black curve in Fig.\ref{fig.rho_0.1r200}, see also discussion in
Sect.2.1).\\
Given the temperature behaviour of $\rho_g(0.1 r_{200})$ and the value of the IC gas
temperature at the same radius, the $S-T$ relation for magnetized clusters predicted by
our model can be calculated using eq.(\ref{eq.entropy}) and is shown in Fig.\ref{fig.st}.
We stress that, based on the HE condition for the IC gas, one has to expect an entropy
scaling with the gas temperature which is different from the one, $S \sim T$, usually
associated with "standard self-similar" models, even in the case $B=0$; the actual
scaling is, in particular, steeper than  $S \sim T$ due to the decreasing behaviour of
$\rho_g(0.1 r_{200})$ with increasing $T$ (see Fig.\ref{fig.rho_0.1r200}).\\
The effect of a fixed amplitude of the magnetic field $B_*$ (set constant here for any
cluster) is to flatten the $S-T$ relation due to the combination of the decrease in $T_g$
caused by the MVT and to the decrease of the IC gas density $\rho_g(0.1 r_{200})$ caused
by the HE condition. The entropy inversion shown for low-$T$ (at fixed value of $B_*$) in
Fig.\ref{fig.st} reflects the sharp cutoff at low-$T$ in the gas density $\rho_g(0.1
r_{200})$ shown in Fig.\ref{fig.rho_0.1r200}.
The binned entropy data derived from Ponman et al. (2003) shown in Fig.\ref{fig.st} are
enclosed by the two curves for $B_* = 2.5$ and $7$ $\mu$G. This indicates that a magnetic
field whose amplitude decreases with decreasing cluster temperature (mass) as $B_* \sim
T_{200}^{\eta}$ with $\eta \approx 0.5$ may reproduce, on average, the whole distribution
of the data. The best-fit trend of the entropy $S(T)$ evaluated by assuming this
phenomenological scaling of the magnetic field amplitude, $B_* \sim T^{0.5}$, is shown in
Fig.\ref{fig.st_fit}.
We also report in Fig.\ref{fig.bt_fit} the best-fit relation for the amplitude of the
magnetic field, $B_* \propto T_{200}^{0.5}$, plus its variation allowed by the IC gas
entropy data.

\begin{figure}[!h]
\begin{center}
 \epsfig{file=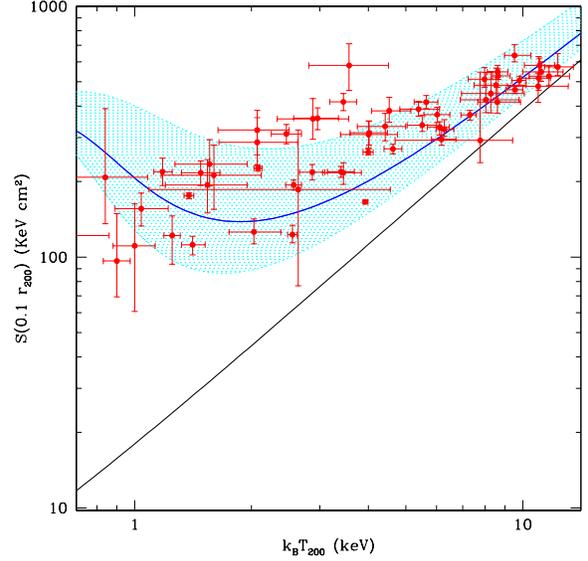,height=8.cm,width=8.cm,angle=0.0}
  \caption{\footnotesize{Same as Fig.\ref{fig.st_fit} but for our best-fit model
  compared to the unbinned data distribution of Ponman et al. (2003).
  }}
  \label{fig.st_all}
\end{center}
\end{figure}

\begin{figure}[!h]
\begin{center}
 \epsfig{file=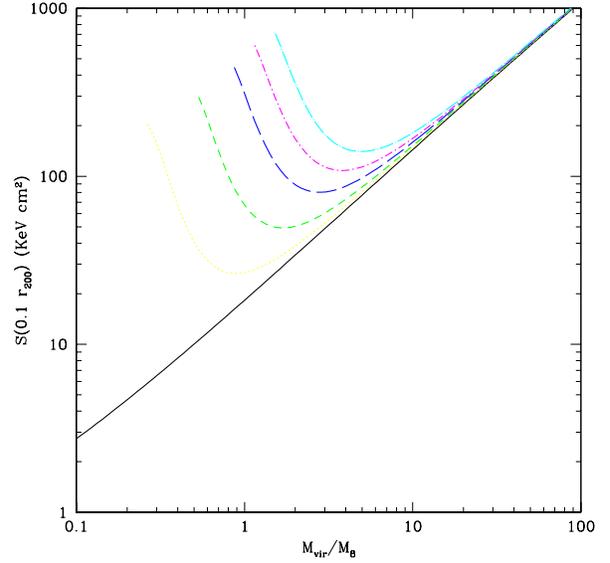,height=8.cm,width=8.cm,angle=0.0}
  \caption{\footnotesize{The entropy $S(0.1 r_{200})$ as a function of the
  cluster virial mass $M_{vir}$ (in units of $M_8$) for various values of the magnetic
  field $B_*$ as in Fig.\ref{fig.st}.
  }}
  \label{fig.sm}
\end{center}
\end{figure}

We notice that the effect of a magnetized IC gas produces an inversion of the entropy at
low temperature values, i.e. in the group region, as the one shown by the unbinned data
derived from Ponman et al. (2003). Fig.\ref{fig.st_all} shows our best-fit prediction
plus the uncertainty region for the $S-T$ relation compared to the unbinned entropy data
of Ponman et al. (2003, see their Fig.4).
We also plot in Fig.\ref{fig.sm} the $S-M$ relation predicted by our model.


\section{The $L_X-T$ relation for magnetized clusters}
\label{sect.lxt}

\begin{figure}[!h]
\begin{center}
 \epsfig{file=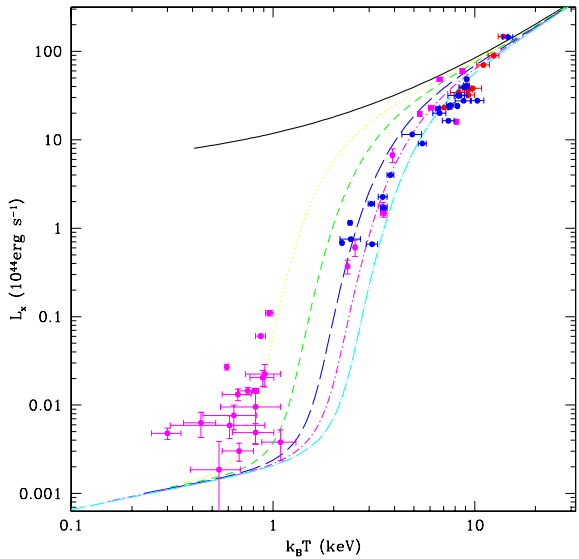,height=8.cm,width=8.cm,angle=0.0}
  \caption{\footnotesize{The bolometric X-ray luminosity $L_X$ as a function of the
  gas temperature for various values of the magnetic field $B_*=0$ (black curve), $2.5$ $\mu$G
  (yellow), $3.7$ $\mu$G (green), $5$ $\mu$G (blue), $6$ $\mu$G (red) and $7$ $\mu$G (cyan).
  Data are taken from Arnaud \& Evrard (1999, blue dots), Allen et al. (2001, red dots)
  and Ponman et al. (2003, magenta dots).
  }}
  \label{fig.lx_t}
\end{center}
\end{figure}

\begin{figure}[!h]
\begin{center}
 \epsfig{file=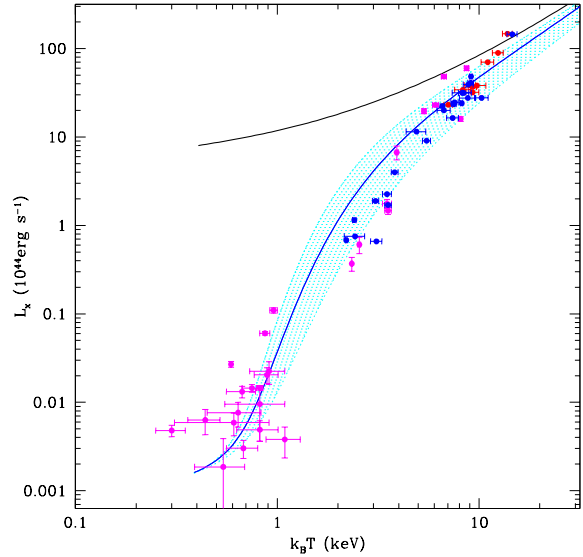,height=8.cm,width=8.cm,angle=0.0}
  \caption{\footnotesize{The bolometric X-ray luminosity $L_X$ as a function of the gas
  temperature for a magnetic field scaling $B_* \sim T_{200}^{\eta}$ with $\eta= 0.5 \pm 0.1$.
  Note that this magnetic field scaling is the same that fits the $S-T$ relation, as shown in
  Fig.\ref{fig.st_fit}.
  }}
  \label{fig.lx_t_fit}
\end{center}
\end{figure}

We show in this section that the same physical model that is able to recover the
flattening of the $S-T$ relation also provides an explanation for the steepening of the
$L_X-T$ relation, from the scale of clusters down to the group scale.\\
Since the X-ray emissivity is mainly provided by thermal bremsstrahlung emission
 \be
 \varepsilon_{brem} \approx 3 \cdot 10^{-27} {\rm erg s^{-1} cm^{-3} }
 \rho_g^2(r,B) T_g^{1/2}(B)\; ,
 \label{eq.xemissivity}
 \ee
(here we adopt its expression for $k_B T_g \simgt 2$ keV, see Sarazin 1988), the
bolometric X-ray luminosity of a cluster with mass $M$ is given by
 \be
 L_X \propto  T_g^{1/2}(B) \rho_g^2(0,B) \bigg( {r_{vir} \over c}\bigg)^3 \int_0^c dx
 x^2 y_g(x,B)
 \label{eq.lxbol}
 \ee
which reads, in practical units, as
 \begin{eqnarray}
 L_X & \approx & 2.8 \cdot 10^{45} {\rm erg s^{-1}} \bigg({k_B T_g(B) \over 4 keV}\bigg)^{1/2}
 \bigg({n_g(0,B) \over 10^{-3} cm^{-3}} \bigg)^2 \times \nonumber \\
     & & {1 \over c^3} \bigg( {M_{vir} \over M_8}\bigg) \int_0^c dx x^2 y_g(x,B) \, .
 \label{eq.lxbol_units}
 \end{eqnarray}
The presence of a magnetic field provides a strong decrease of the cluster $L_X$ at low
temperature due to the decrease of the central IC gas density (see Sect.2) and also a
decrease of the cluster temperature from the MVT with respect to their unmagnetized
values.\\
The result of a simple calculation based on the HE, on the MVT and on the expression of
$L_X$ as given in eqs.(\ref{eq.lxbol},\ref{eq.lxbol_units}) is reported  in
Fig.\ref{fig.lx_t} and it shows that the effect of the magnetic field is to steepen the
$L_X-T$ relation.
We stress that the X-ray luminosity data shown in Fig.\ref{fig.lx_t} are enclosed by the
same curves that enclose the $S-T$ data in the same temperature range. Moreover, as in
the $S-T$ relation, a magnetic field whose amplitude decreases with decreasing cluster
mass as $B_* \sim T_g^{0.5 \pm 0.1}$ is able to reproduce the whole distribution of the
$L_X-T$ data from groups to clusters. The best-fit trend of the $L_X-T$ relation
evaluated by assuming this phenomenological scaling of the magnetic field amplitude which
fits the entropy data (see Figs. 8 and 10) is shown in Fig.\ref{fig.lx_t_fit}.

\begin{figure}[!h]
\begin{center}
 \epsfig{file=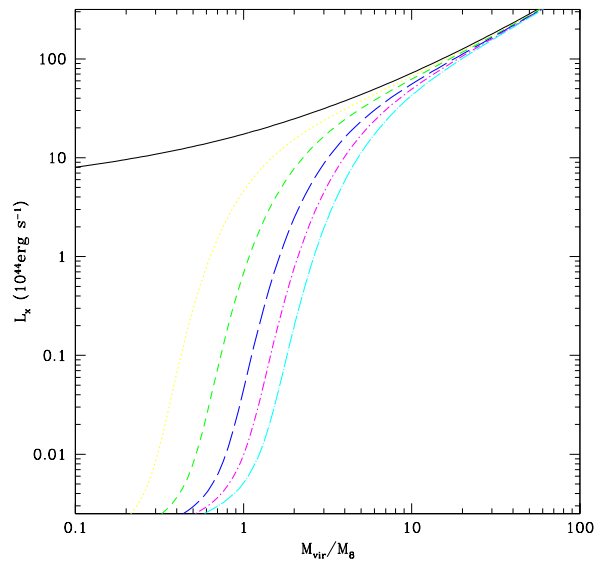,height=8.cm,width=8.cm,angle=0.0}
  \caption{\footnotesize{The bolometric X-ray luminosity $L_X$ as a function of the
  cluster mass for various values of the magnetic field $B_*=0$ (black curve), $2.5$ $\mu$G
  (yellow), $3.7$ $\mu$G (green), $5$ $\mu$G (blue), $6$ $\mu$G (red) and $7$ $\mu$G (cyan).
  }}
  \label{fig.lx_m}
\end{center}
\end{figure}

We also show in Fig.\ref{fig.lx_m} the $L_X-M$ relation that is consistent with the
$L_X-T$ relation previously discussed.

The fact that the same model for magnetized clusters which is able to reproduce the $S-T$
relation (see Sect.5) also recovers the $L_X-T$ relation is not a coincidence, but is a
consequence of the (reliable) physical description of the X-ray cluster structure. In
fact, the entropy $S = T_g/ \rho_g^{2/3}$ calculated at a scale $0.1 r_{200}$ (i.e.,
within the cluster core) sets the values of the gas density and temperature at the same
spatial scale, and hence the value of the X-ray luminosity $L_X \propto \rho^2_g T_g$,
which is mostly determined by the cluster core properties. Therefore, adjusting the
parameters of our model to match the observed $S-T$ relation implicitly ensures that the
resulting $L_X-T$ relation is close to the observed one.


\section{Exploring the effects of boundary conditions}
 \label{sect.boundary}

In the previous sections we have shown the effect of the magnetic field on the $S-T$ and
$L_X-T$ relations for the case in which i) the external pressure was negligible (the
formal case $P_{ext} = 0$ has been adopted in Sects. 4-6) and ii) the boundary condition
used to normalize the IC gas density profile given by eq.(\ref{eq.rho_norm}), i.e. a
constant gas mass within the cluster virial radius.

For completeness, we describe in this section the effects on cluster temperature, entropy
and X-ray luminosity induced by changing the previous boundary conditions.

\subsection{Normalizing to the cluster virial pressure}
 \label{sect.pextconst}

\begin{figure*}[!t]
\begin{center}
\hbox{
 \epsfig{file=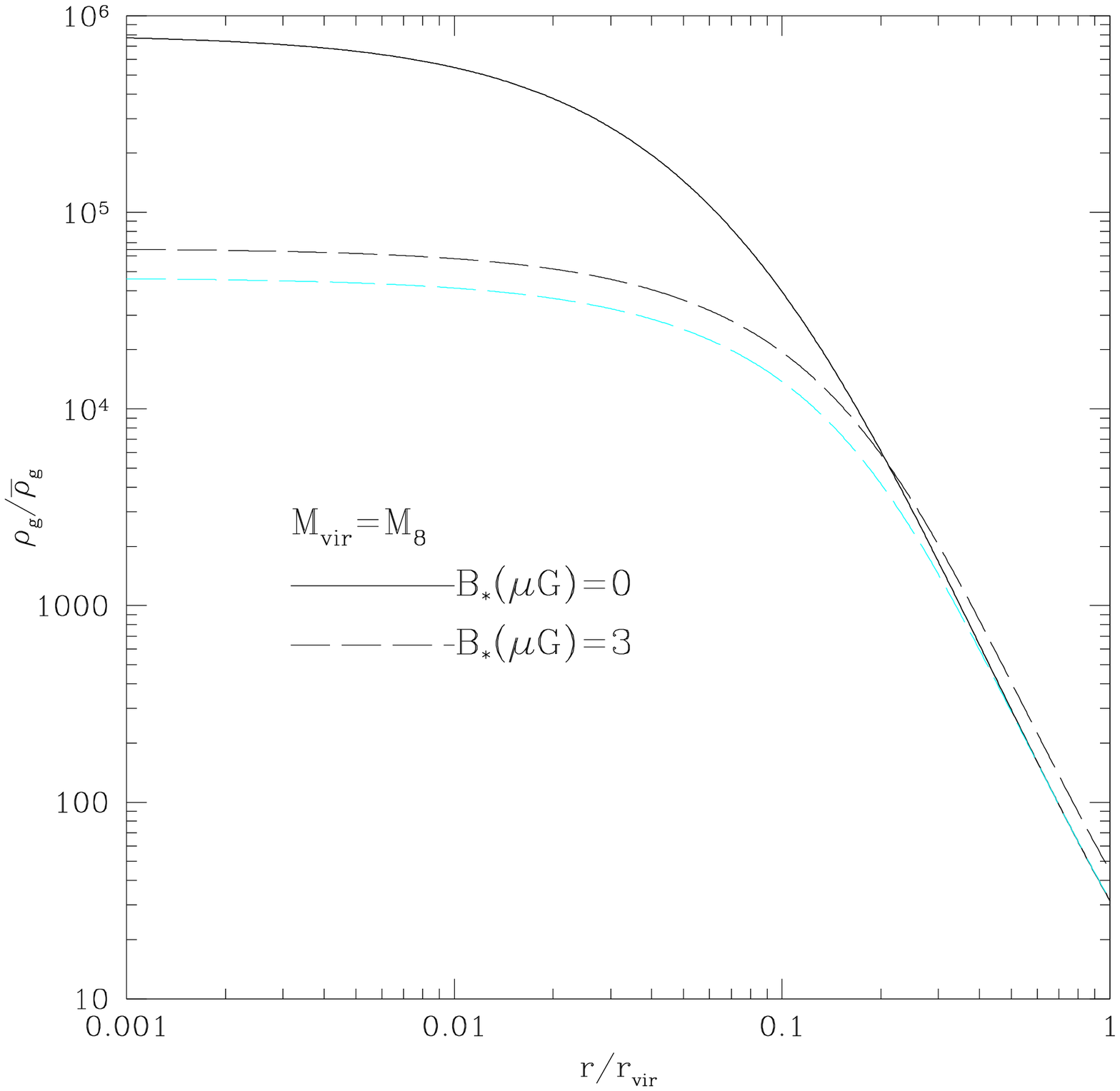,height=8.cm,width=8.cm,angle=0.0}
 \epsfig{file=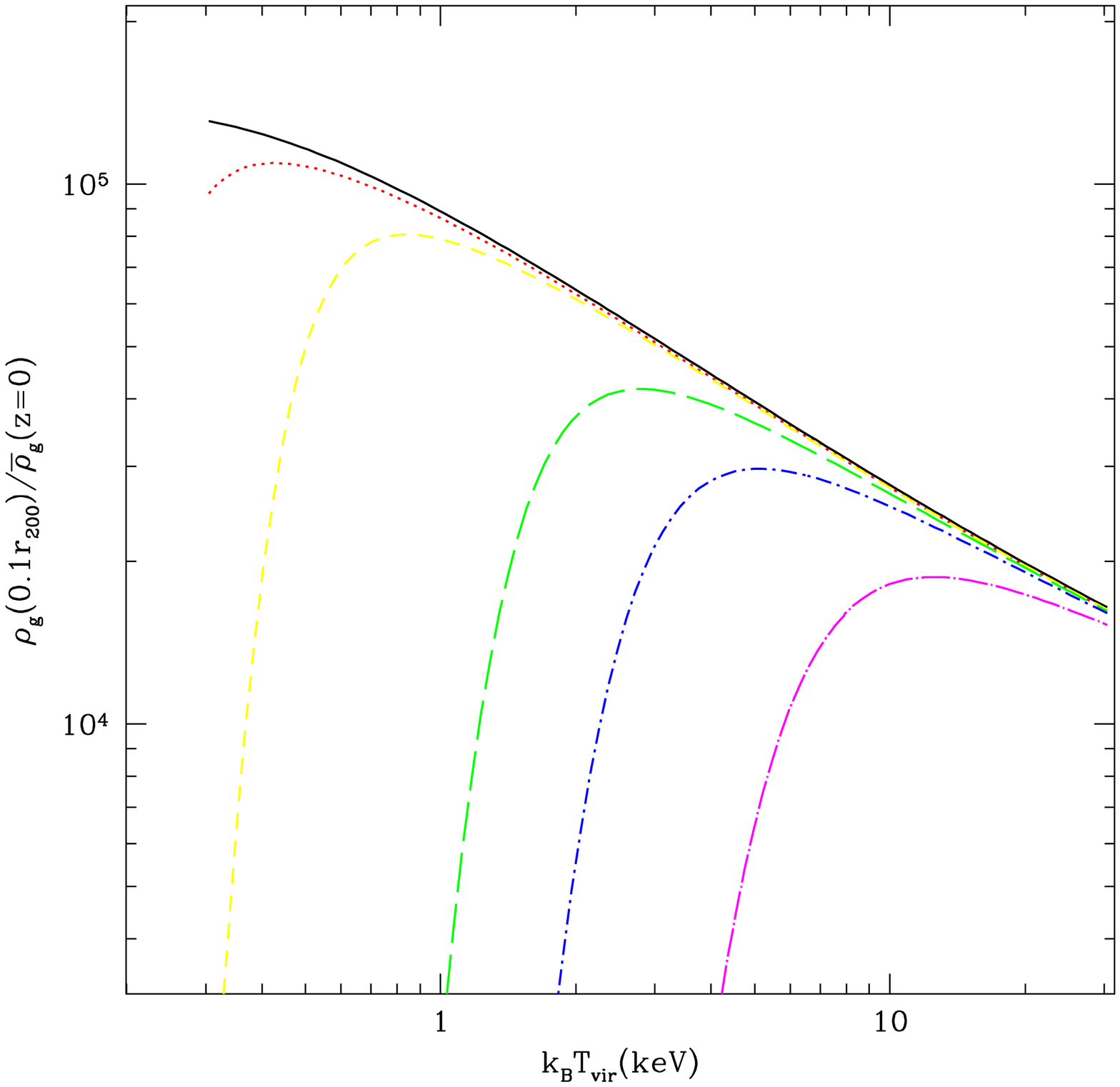,height=8.cm,width=8.cm,angle=0.0}
 }
\end{center}
  \caption{\footnotesize{{\bf Left}. The radial density profiles of the IC gas normalized to the
  same virial pressure $P_{vir}$ for values of the magnetic field $B_*=0$ (solid) and
  $B_*=3$ $\mu$G (dashed cyan) are compared to the analogous density profiles normalized to
  the total gas mass at $r_{vir}$ (dashed black). A cluster with $M= M_8$ is considered here.
  {\bf Right}. The temperature dependence of the IC gas density evaluated at $0.1 r_{200}$ for
  the same values of the B-field as in Fig.\ref{fig.rho_0.1r200} but for a normalization of the
  IC gas density to the virial pressure $P_{vir}$.
  }}
  \label{fig.rho_pvir}
\end{figure*}

\begin{figure*}[!t]
\begin{center}
\hbox{
 \epsfig{file=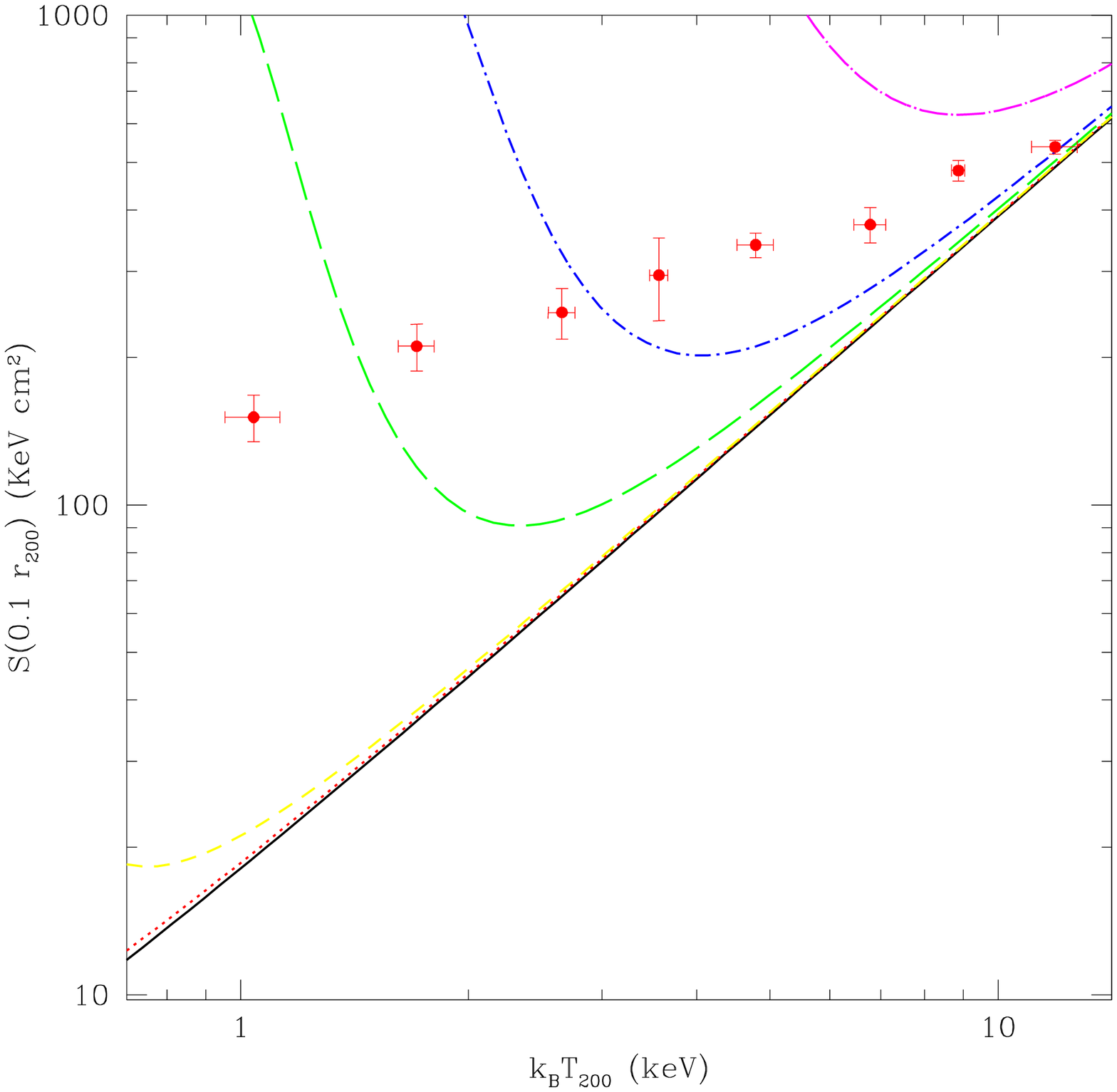,height=8.cm,width=8.cm,angle=0.0}
 \epsfig{file=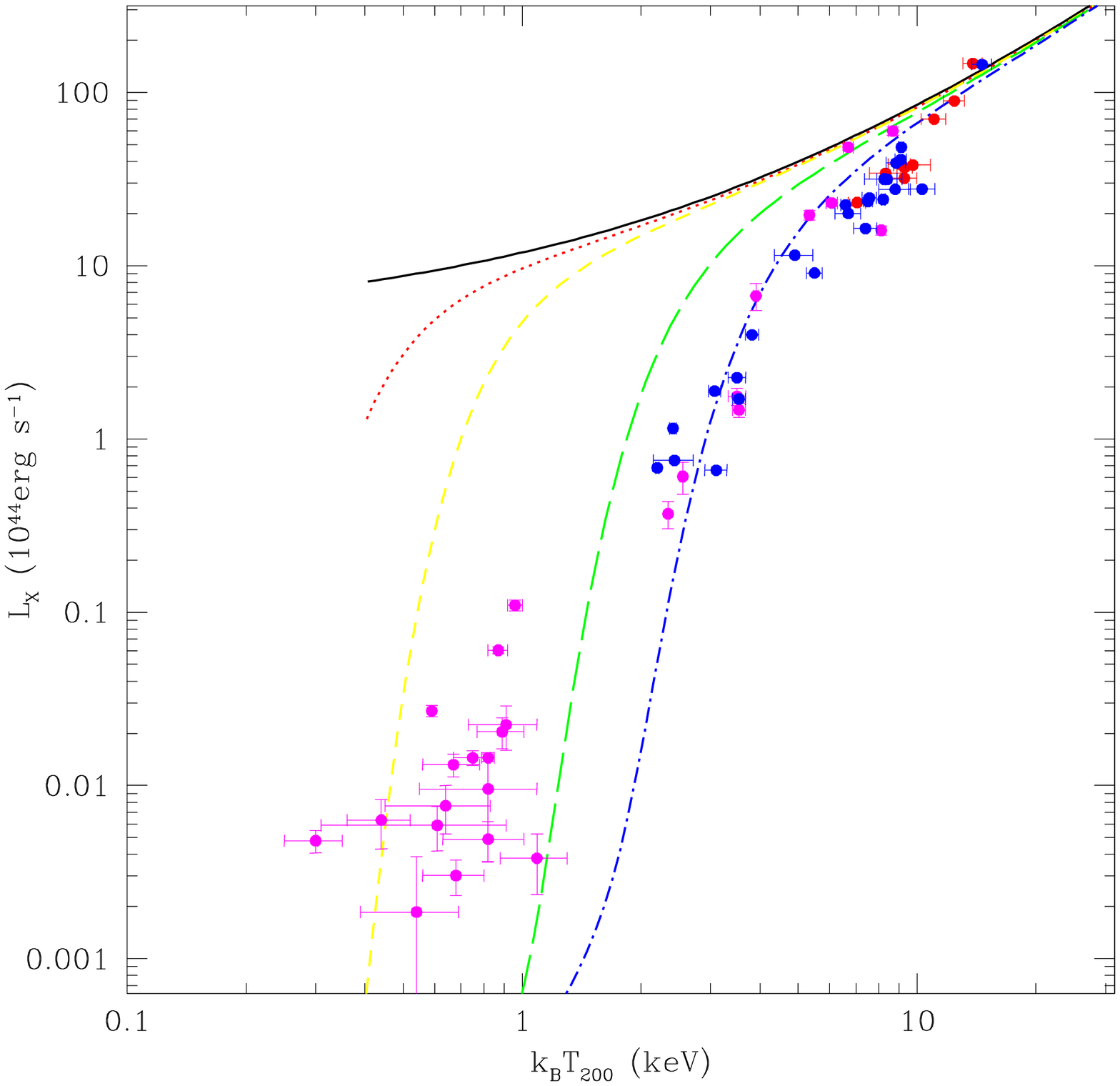,height=8.cm,width=8.cm,angle=0.0}
 }
\end{center}
  \caption{\footnotesize{{\bf Left}. The $S-T$ relation for clusters with IC gas density
  profiles normalized to the virial pressure $P_{vir}$ is shown for the same values of the
  magnetic field $B_*$ as in Fig.\ref{fig.st}.
  {\bf Right}. The $L_X-T$ relation for clusters with IC gas density
  profiles normalized to the virial pressure $P_{vir}$ is shown for the same values of the
  magnetic field $B_*$ as in Fig.\ref{fig.lx_t}.
  }}
  \label{fig.ST_LXT_pvir}
\end{figure*}

Normalizing the IC gas density profile to have a constant pressure at the virial radius,
$P_g(r_{vir},B)=$ const., yields a decrease of the central gas density with increasing
values of $B_*$ which is larger than when normalizing the same density profile to
$M_{g}(r_{vir})$; this happens because, for a given cluster mass (temperature), the
former condition requires $\rho_g(r_{vir},B)= \rho_g(r_{vir},B=0)$. This fact implies a
behaviour of the function $K(M,B) = \rho_g(r,B)/\rho_g(r,B=0)$ that is slightly different
from that shown in Fig.\ref{fig.kappa}, and consequently a slightly different behaviour
of the scaling of $\rho_g(0.1 r_{200})$ with the cluster temperature (mass) as shown in
Fig.\ref{fig.rho_pvir}.
Since the maxima of the function $\rho_g(0.1 r_{200})$ are moved to lower temperatures in
this case, the inversion of the $S-T$ relation is also moved at lower temperatures and
the flattening of the $L_X-T$ relation is found at lower values of the X-ray luminosity
and lower temperatures (see Fig.\ref{fig.ST_LXT_pvir}).
These results show that both the entropy inversion and the flattening of the $L_X-T$
relation at low temperatures are a realistic effect to be expected in models of
magnetized clusters; moreover, these results do not strongly depend  on the adopted
boundary conditions for the gas density profile of the clusters. This reinforces our
conclusions on the ability of models of magnetized clusters to reproduce the basic
structural properties of X-ray clusters.\\
To conclude our exploration of the effects of boundary conditions, we will address in the
next section the effects on gas density, entropy and X-ray luminosity caused by changing
the external pressure $P_{ext}$ at the cluster boundary.

\subsection{Changing the cluster external pressure}
 \label{sect.pext}

\begin{figure*}[!t]
\begin{center}
\hbox{
 \epsfig{file=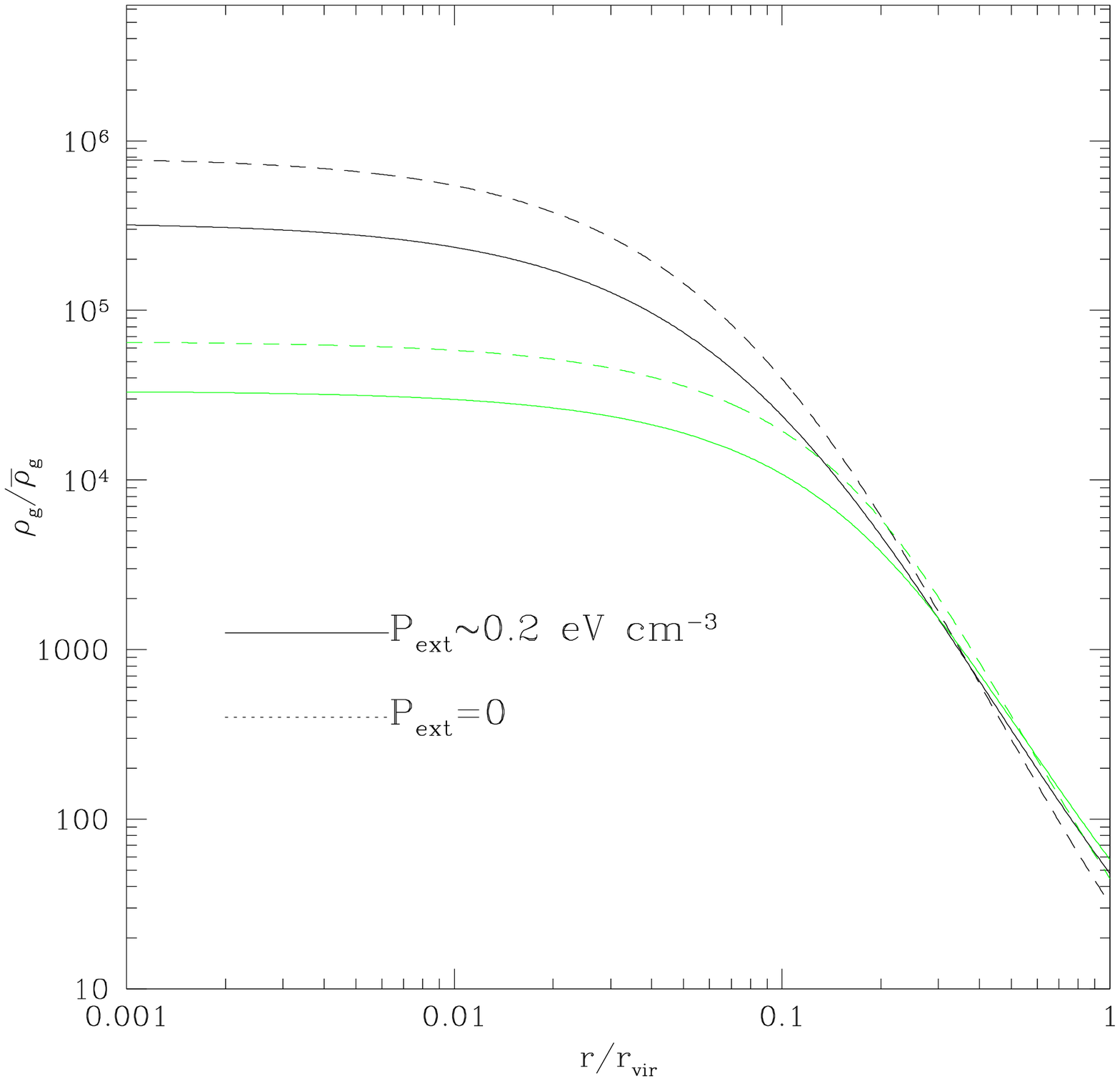,height=8.cm,width=8.cm,angle=0.0}
 \epsfig{file=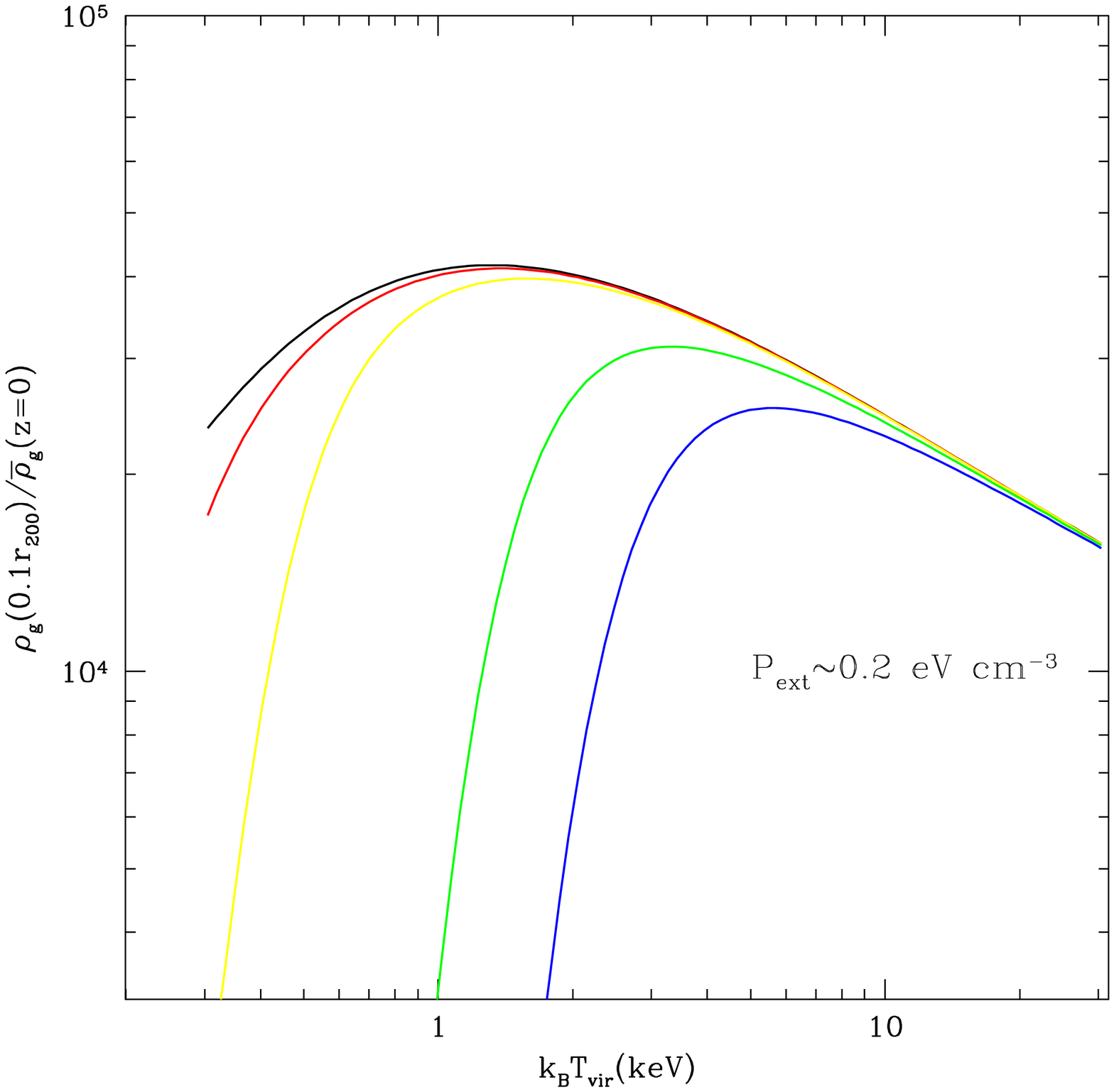,height=8.cm,width=8.cm,angle=0.0}
 }
\end{center}
  \caption{\footnotesize{{\bf Left}. The radial density profiles of the IC gas evaluated
  in the presence of an external pressure $P_{ext}=0.2$ eV cm$^{-3}$ (solid curves) are shown
  for values of $B_*=0$ (black) and $B_*=3$ $\mu$G (green) and are compared to the
  analogous profiles evaluated with $P_{ext}=0$ (dashed curves for the same magnetic field values).
  A cluster with $M= M_8$ has been chosen.
  {\bf Right}. The temperature dependence of the IC gas density evaluated at $0.1 r_{200}$
  is shown for the same values of the B-field as in Fig.\ref{fig.rho_0.1r200}
  but for a value of the external pressure $P_{ext}=0.2$ eV cm$^{-3}$.
  }}
  \label{fig.rho_pext}
\end{figure*}

\begin{figure*}[!t]
\begin{center}
\hbox{
 \epsfig{file=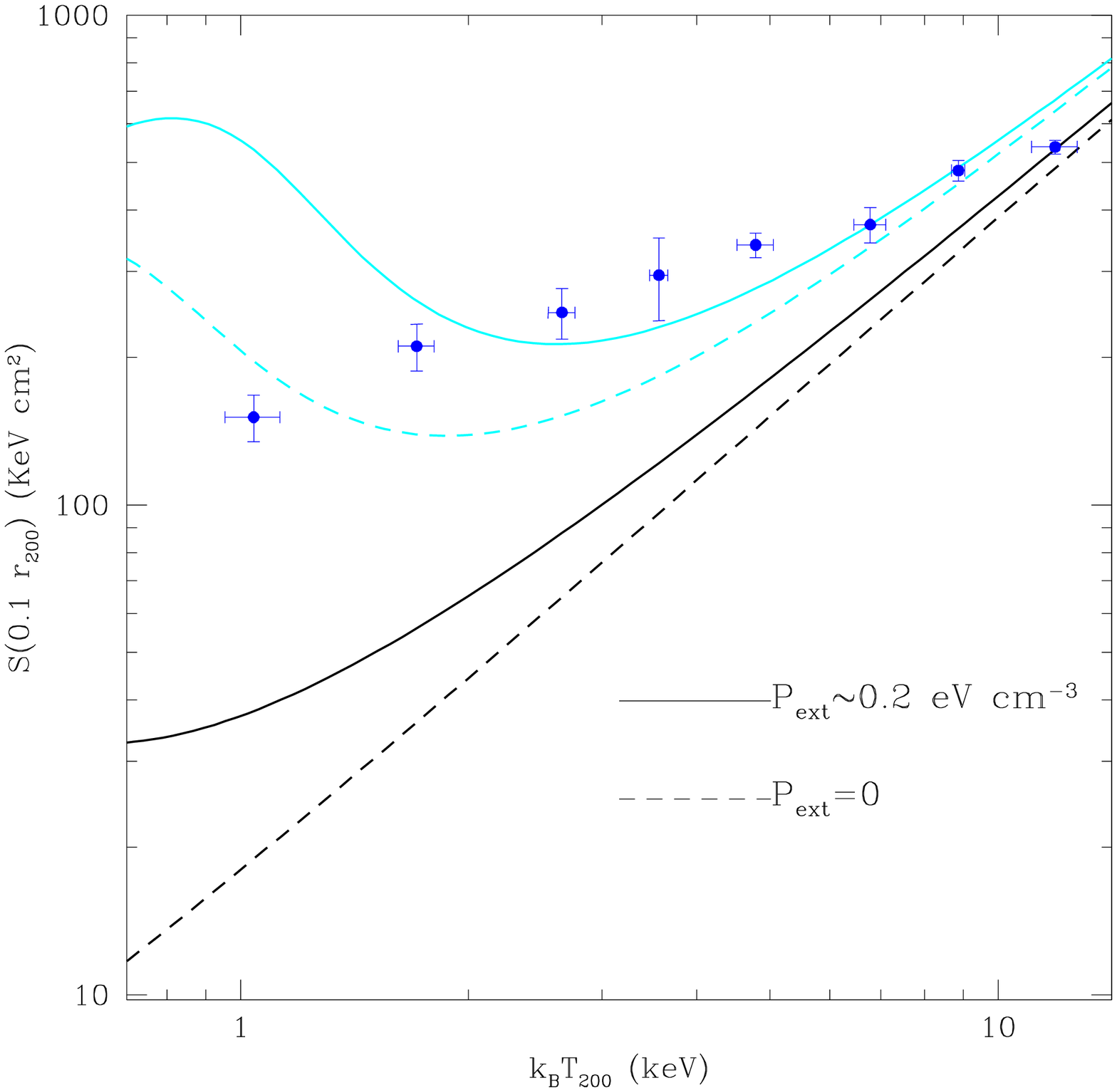,height=8.cm,width=8.cm,angle=0.0}
 \epsfig{file=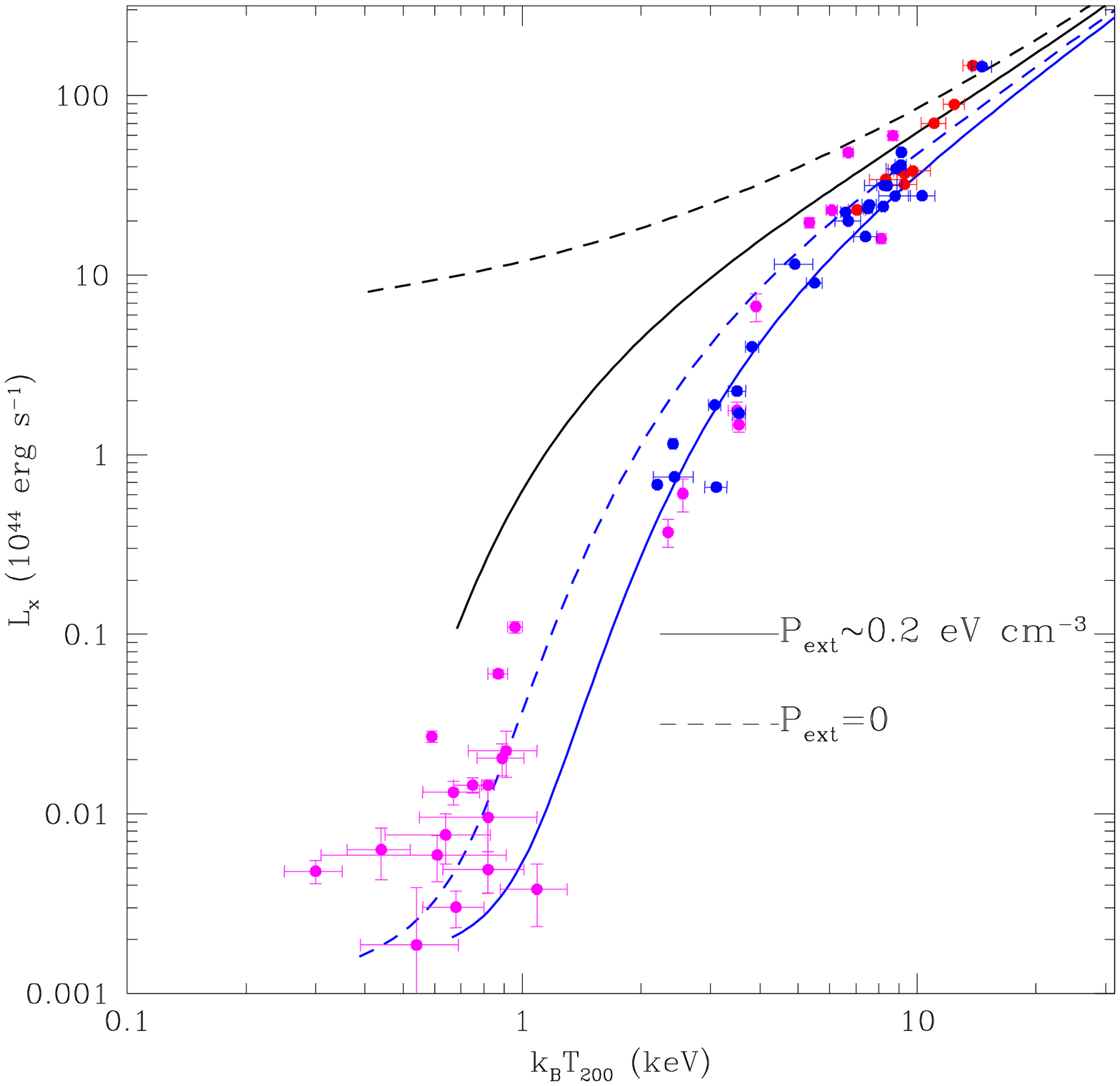,height=8.cm,width=8.cm,angle=0.0}
 }
\end{center}
  \caption{\footnotesize{{\bf Left}. Comparison of the best fit $S-T$ relation
  evaluated for $P_{ext}=0$ (dashed cyan, see Fig.\ref{fig.st_fit})
  and the one calculated for the same parameters but for $P_{ext}=0.2$ eV cm$^{-3}$ (solid cyan).
  The $S-T$ relations evaluated in the case $B_*=0$ with $P_{ext}=0$ (dashed black) and $B_*=0$
  with $P_{ext}=0.2$ eV cm$^{-3}$ (solid black) are also shown for comparison.
  {\bf Right}. The same cases shown for the $S-T$ relation in the left panel
  of this figure are shown here for the $L_X - T$ relation.
  Different curves refer to the same cases as discussed in the left panel.
  }}
  \label{fig.ST_LXT_pext}
\end{figure*}

Changing the value of the external pressure $P_{ext}$ in the MVT modifies the cluster
temperature and also alters the solution $\rho_g(r,B)$ of the HE condition as shown in
Fig.\ref{fig.rho_pext}. In fact, increasing the value of $P_{ext}$ yields an increase of
the overall cluster temperature (see eq.\ref{eq.TvirM}) and hence a decrease of the
central IC gas density, to be consistent with its boundary conditions (see
Fig.\ref{fig.rho_pext}).
The increase of $P_{ext}$ in the MVT and in the HE condition causes a larger decrease of
the IC gas density in the cluster core (even in the absence of a magnetic field) and
hence a further bending of the temperature dependence of $\rho_g(0.1 r_{200})$ in the
low-T region than the case in which $P_{ext}=0$ (see Fig.\ref{fig.rho_0.1r200}).
This behaviour implies that the $S-T$ relation is flatter at low-T with respect to the
case in which $P_{ext}=0$, even for $B_*=0$ (see Fig.\ref{fig.ST_LXT_pext}). In addition,
the entropy inversion which is produced in the low-T region by the effect of magnetic
field is further enhanced by the increasing value of $P_{ext}$.
As a direct consequence of this fact, the $L_X-T$ relation computed for values $P_{ext}
>0 $ becomes steeper than that derived in the case $P_{ext}=0$, even in the absence of a
magnetic field. The presence of a magnetic field (like the one $B_* \propto T^{0.5}$ that
best fits the $S-T$ and the $L_X-T$ relations for $P_{ext}=0$, see Figs.\ref{fig.st_fit}
and \ref{fig.lx_t_fit}) further increases the steepness of the predicted $L_X-T$ relation
towards low temperatures, thus requiring lower values of the magnetic field to reproduce,
in this case,  the distribution of the available data  (see Fig.\ref{fig.ST_LXT_pext}).

\begin{figure*}[!t]
\begin{center}
\hbox{
 \epsfig{file=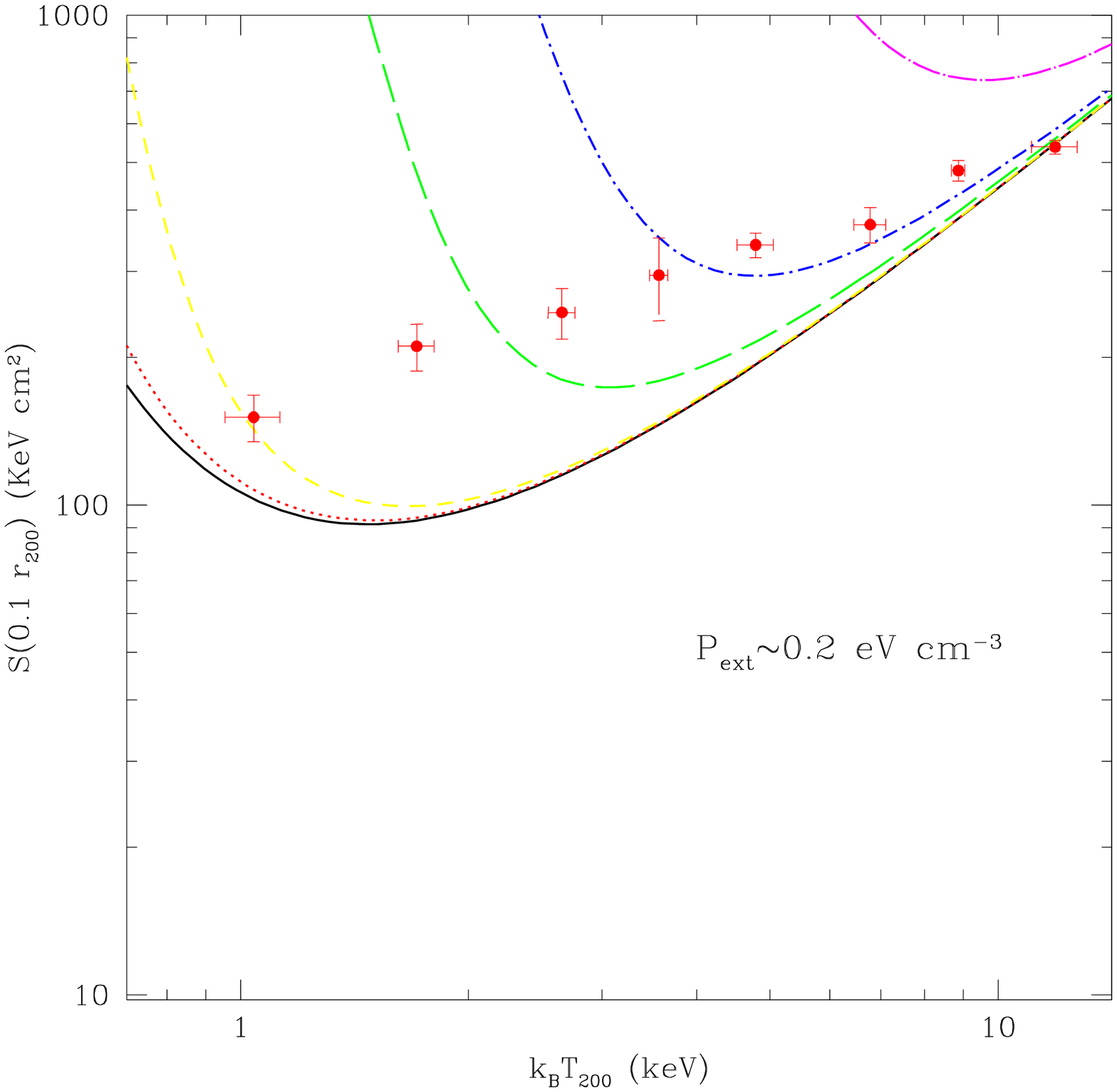,height=8.cm,width=8.cm,angle=0.0}
 \epsfig{file=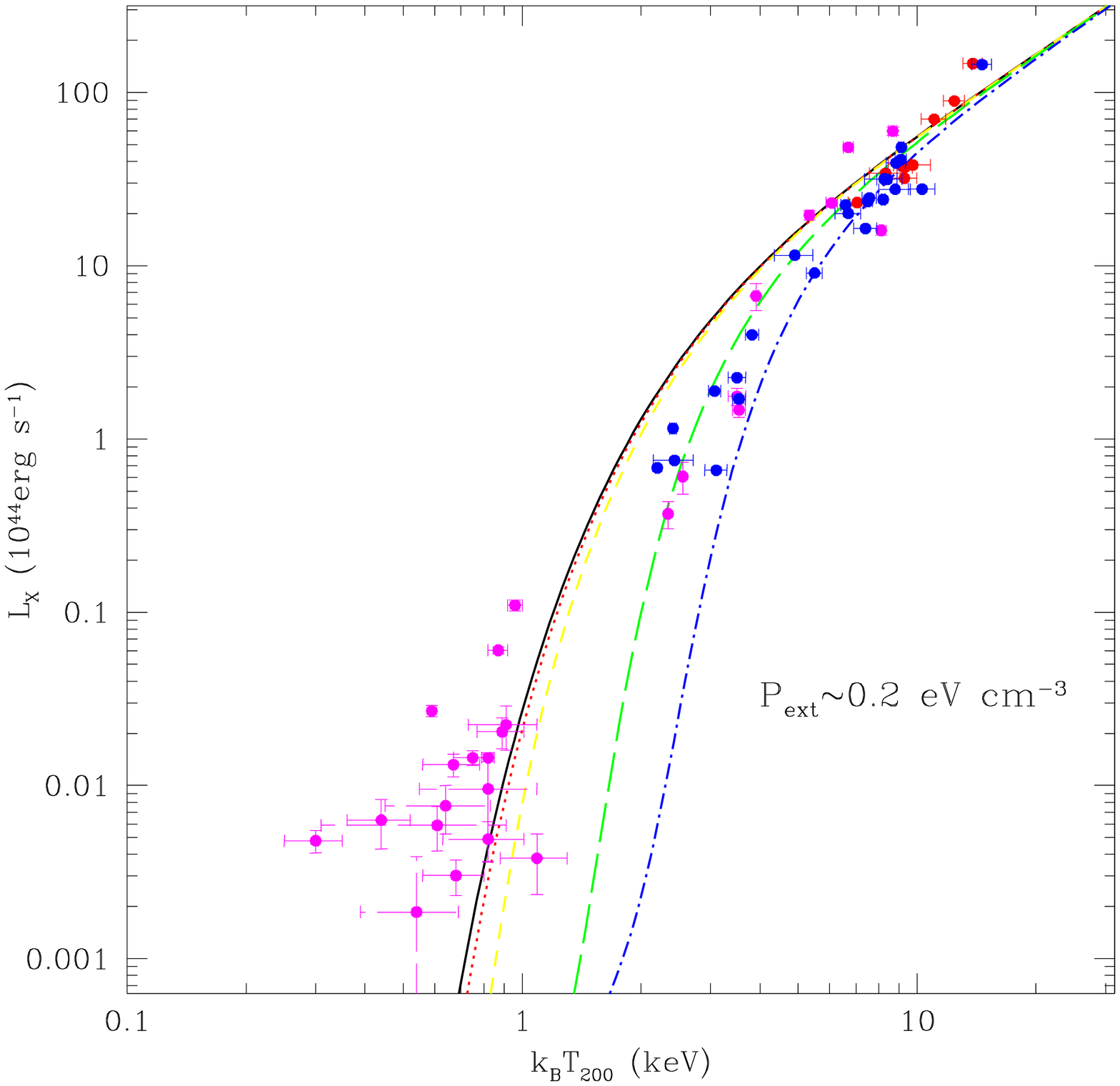,height=8.cm,width=8.cm,angle=0.0}
 }
\end{center}
  \caption{\footnotesize{{\bf Left}. The $S-T$ relation for clusters with IC gas
  density profiles normalized to the virial pressure $P_{vir}$ and with $P_{ext}=0.2$ eV cm$^{-3}$.
  Curves are for the same values of the magnetic field $B_*$ as in Fig.\ref{fig.st}.
  {\bf Right}. The $L_X-T$ relation for clusters with IC gas density
  profiles normalized to the virial pressure $P_{vir}$ and with $P_{ext}=0.2$ eV cm$^{-3}$.
  Curves are for the same values of the magnetic field $B_*$ as in Fig.\ref{fig.lx_t}.
  }}
  \label{fig.ST_LXT_pvir_pext}
\end{figure*}

Finally, the combination of large values of $P_{ext}$ and a different boundary condition
(like the one requiring us to normalize the IC gas density profile to the virial
pressure, that we previously explored in this section) has the combined effect of further
increasing both the flatness and the inversion of the $S-T$ relation and, consistently,
the steepness of the $L_X-T$ relation in the low-temperature region (see
Fig.\ref{fig.ST_LXT_pvir_pext}).
We stress that in this last case the $S-T$ and the $L_X-T$ relations evaluated for a null
magnetic field are no longer representable by simple power-law behaviours, since the
presence of a positive external pressure alters the cluster temperature derived from the
MVT and the value of $\rho_g(0.1 r_{200})$.
This suggests that, in this case, the observed $L_X-T$ relation (as well as the S-T
relation) could be roughly reproduced by the combination of high values of $P_{ext}$ with
a more modest contribution from the magnetic field (with respect to the case
$P_{ext}=0$), because the effect of $P_{ext}$ in the MVT and in the HE conditions of our
model (see Sect.2) tends to mimic -- at a given boundary condition for the IC gas at the
cluster virial radius -- that of an increasing magnetic field on the IC gas density
profile. In fact, the external pressure compresses the IC gas and tends to rise its
temperature, thus requiring a decrease of the core density in order to fulfill the IC gas
boundary condition at $r_{vir}$. Both the boundary conditions that have been considered
here ($M_g(r_{vir})=$ const., or $P_g(r_{vir})=$ const.) provide hence a decrease of the
IC gas density in the cluster core which affects both the entropy and the X-ray
luminosity, as we have shown here.\\
By stretching our analysis to the extreme conclusion, we could say that the $S-T$ and the
$L_X-T$ relations could be reproduced with an appropriate distribution of external
pressure values, decreasing from the rich cluster scale down to the group scale, with
only a small effect due to the unavoidable presence of a low-amplitude intra-cluster
magnetic field.
However, we recall that values of $P_{ext} \approx 0.2$ eV cm$^{-3}$ (considered in
Figs.\ref{fig.rho_pext} - \ref{fig.ST_LXT_pvir_pext}) are found at the outskirts ($r
\simgt r_{vir}$) of rich, hot clusters (with mean temperatures $\simgt 10$ keV, see
Sect.2.2) and they are, therefore, extreme values of the external pressure which are much
higher than those usually found around low-T clusters and groups. This means that the
temperature scaling of the cluster entropy and X-ray luminosity (with and without a
magnetic field) for such a high value of $P_{ext}$ considered as representative of the
whole population of clusters (from rich ones down to small groups) has to be considered
unlikely; a lower value of $P_{ext}$ closer to the value found in the IGM is better
suited to the case of low-T systems (see, e.g., our discussion in Sect.2, see also the
results of the temperature and density profile analysis of a sample of poor and rich
clusters studied by Piffaretti et al. 2005), and thus the results presented in Sects. 3-6
seem to be more reliable.

The results discussed in this Section have been presented for illustrative purposes and
are based on a parametric variation of the leading physical quantities and boundary
conditions. A more detailed exploration of these effects requires a thorough analysis of
the density and temperature profiles of a large sample of groups and rich clusters out to
radii $r \simgt r_{vir}$, an analysis which is crucial for this issue, but it is not yet
available. The previous analysis should be combined with the complementary study of the
evolution of the intra-cluster magnetic field with the cluster scale (mass, radius) which
has just begun and that will receive a substantial boost from the advent of the next
generation radio observatories with high sensitivity and spatial resolution (e.g, SKA,
LOFAR).
Given the lack of detailed data on these subjects, we will tackle in detail the
theoretical aspects related to these issues in a forthcoming paper (Colafrancesco et al.
2006, in preparation).


\section{Discussion and conclusions}

Our results provide an analytical description of the effects of a magnetic field on the
density structure and on the entropy profiles of galaxy clusters in the context of the HE
condition and of the MVT. We have shown that such a description is able to provide, at
once, a possible explanation of three problematic aspects of the cluster structure:
i) the flattening of the entropy profile in the cluster center (together with the
possible entropy inversion near the center of the structure);
ii) the flatness of the $S-T$ relation (together with the entropy inversion at the scale
of the galaxy groups);
iii) the increasing steepening of the $L_X-T$ relation from the cluster scale towards the
group scales.

The inclusion of the effects of the magnetic field in the description of the
intra-cluster gas structure is motivated by the increasing evidence for the presence of a
magnetic field in these cosmic structures (see, e.g., Carilli and Taylor 2002, Govoni and
Feretti 2005 for recent reviews).

In the framework of our model, all the available data indicate that an increase of the
magnetic field $B_* \sim T^{0.5 \pm 0.1}$ is able to reproduce, at the same time, both
the $S-T$ and the $L_X-T$ relations. Such a scaling is a reasonable expectation of
hierarchical clustering scenarios in which the magnetic field is coupled to the amount of
gas collected in the deepening potential wells of increasing mass systems.
The available evidence on the amplitudes of cluster magnetic fields for individual
clusters and groups seem to be consistent with this result (see also Cassano et al. 2006
for an independent study).\\
Since the theoretical results presented here partly depend on the assumed boundary
conditions for the IC gas at $r \approx r_{vir}$, we have also provided a systematic
exploration of the effects of changing the IC gas boundary conditions on the structure
and evolution of X-ray clusters. Such an analysis has been able to show that the study of
the magnetic field effects on the IC gas must be addressed in full complementarity with
the pressure and density study of the cluster external regions at $r \simgt r_{vir}$.

How does the model and the results we have presented here compare to other theoretical
explorations of the IC gas physics?\\
It has been suggested (see, e.g., Arnaud 2005 for a review) that the steepening of the
$L_X-T$ relation could be due to a systematic change (increase) of the mean gas density
with $T$ (i.e., a simple modification of the scaling laws) or to a variation of the
cluster shape with $T$ (i.e., a breaking of self-similarity). Many studies have
attributed the deviations from self-similarity to an early episode of preheating by
supernovae or active galaxies so that the heat input that estabilishes a uniform minimum
entropy level in the intergalactic medium breaks the self-similarity because the extra
entropy makes the gas harder to compress as it accretes into dark matter halos (e.g.,
Kaiser 1991, Evrard \& Henry 1991). However, several analyses suggested that the energy
input required to explain the observed relations through global preheating seems to be
quite extreme (see, e.g., Donahue et al. 2006, Arnaud 2005).
Alternative explanations (Voit \& Bryan 2001) have argued that the entropy scale
responsible for the breaking of self-similarity is not a global property of the IGM but
rather a condition set by the radiative cooling. In such a scenario, cooling and feedback
seem to conspire to deplete the amount of gas below the cooling threshold: thus, some of
the low-entropy gas condenses and the feedback subsequently raises the entropy of the
remaining gas until both cooling and feedback shut down. Variations on this theme (see,
e.g., Voit et al. 2002) are also able to provide models that may account for several
X-ray observables (like the $L_X-T$ and the $T-M$ relations, the X-ray surface brightness
profiles of clusters and their temperature gradient). However, these models are not yet
able to explain why clusters differ in the amount of gas that still remains below the
cooling threshold.

In such a complex theoretical framework, some recent studies of the self-similarity in
the cluster gas temperature and entropy profiles suggest that the departures of the
entropy and X-ray luminosity scaling laws from the standard self-similar model could be
due to a change in the temperature (mass) dependence of the normalization of the gas
density profiles (see Arnaud 2005 for a general discussion).\\
In our model a non-standard scaling of the IC gas density with the cluster temperature
(mass) is provided (see Figs.\ref{fig.rho_b_norm} and \ref{fig.rho_0.1r200}) by the
inclusion of the magnetic field effects on the HE (and the MVT) condition of the IC gas,
which then changes both the radial distribution of the IC gas (for a given cluster mass
and B-field value) and the relation between the gas temperature and cluster mass.
It follows that a consistent description of the magnetized ICM can provide a simple
explanation of several (or of all) of these still open problems and thus weakens the need
for the inclusion of non-gravitational effects -- like e.g., shock heating, AGN or SN
feedback -- which have been proposed, so far, for the explanation of these features.

Our study (far from being conclusive) shows that a more global description of the IC gas
physics may help in solving some of the crucial aspects of cluster structure and
evolution.\\
In these respects, we want to stress, in the following, some of the limits of this
initial study.

The isothermal case that we explored here in detail is not valid for all clusters and
groups, even though it seems to be a valid approximation in the intermediate regions of
many clusters, from where the entropy data are taken  (see, e.g., Pratt et al. 2006,
Pointecouteau et al. 2005). Thus, even though our comparison with the entropy data taken
from Pratt et al. (2006) and from Ponman et al. (2003) can be quite safe, we more refined
analyses of both the very inner parts (the cool cores at $r \simlt 0.1 r_{200}$) and the
external regions (those at $r \simgt r_{200}$) of clusters.
The cluster cool cores that show quite high values of the magnetic field, sharp $T$ jumps
and probably not isentropic conditions (see, e.g., Piffaretti et al. 2005) and are also
sites of complex AGN -- IC gas interaction (see, e.g., Fabian 2005), are not considered
in a complete self-consistent way here. Our approach to the description of the magnetized
IC gas in this region needs a refined analysis of both the MVT and of the HE equation
that is beyond the scope of this paper, and will be presented elsewhere.
The external regions at $r \simgt r_{200}$ usually show a temperature decrease towards
the outskirts of the cluster and thus, they also need a further analysis with respect to
the isothermal model. The use of a polytropic model (i.e., an extension of our model that
we have already worked out in the Appendix) will be also presented elsewhere.
The study of the pressure and density structure of the cluster external regions is also
relevant to set the appropriate boundary conditions for the IC gas and therefore it
requires a more detailed analysis.

Within the limits of its applicability, this first exploration indicates, nonetheless,
that a more detailed physical description of the IC gas properties that takes into
account also the magnetic field might solve a number of problematic features in cluster
structure and evolution.\\
The predictions of our model can be subject to observational tests. One of the most
direct tests is to look for the amplitude of the magnetic field in clusters and groups
through FR measurements that could directly prove or disprove the scaling of the magnetic
field amplitude $B_* \sim T^{0.5 \pm 0.1}$ that we predict in order to recover the $S-T$
and $L_X-T$ relations. The few data pints available so far seem to be consistent with
such a scaling (see also Cassano et al. 2006, for an independent argument). Detailed FR
measures in the inner $\sim$ Mpc of groups and clusters will be available in the near
future with the forthcoming SKA and LOFAR radio experiments. Some of the cluster will
also likely show evidence for extended and diffuse radio emission that will provide
further information on the radial distribution of the magnetic field (see, e.g.,
Colafrancesco et al. 2005 for a more general discussion).
Lastly, the combination of high-sensitivity X-ray and radio observations will be able to
shed light on the physics of the intra-cluster gas and its associated magnetic field in
both the cool cores and in the external regions of clusters at different redshifts, so
that it will be possible to determine its detailed physical status as well as its cosmic
evolution.


\acknowledgements{
We thank the Referee for useful comments that led to improve the presentation and the
clarity of our paper.
We thank A.Sanderson and T.Ponman for having provided us with the unbinned entropy data
for clusters and groups shown in Fig.\ref{fig.st_all}.
S.C. acknowledges support by PRIN-MIUR under contract No.2004027755$\_$003.}


\appendix
\section{The hydrostatic equilibrium equation in the presence of B-field.}
\label{appendix}

The radial profile of the IC gas density $\rho_g$ can be obtained by solving the
hydrostatic equilibrium equation once we specify the equation of state of the gas. In
general one writes
\be
\nabla p=-\rho_g\nabla\phi(r),\label{eq.idr.gen} \ee
where $p$ is the sum of all the pressure components in the cluster (IC gas, magnetic
field, etc.) and $\phi(r)$ is the gravitational potential. Under the assumption of
spherical symmetry and local homogeneity of the gas, eq.(\ref{eq.idr.gen}) writes as
\be
\frac{1}{\rho_g}\frac{d p}{d r}=-\frac{d\phi}{dr}=-\frac{GM(r)}{r^2}\label{eq.idr.om}\ee
where $M(r)$ is the total mass enclosed in a sphere of radius $r$. The expression for
$M(r)$ depends on the assumed gravitational potential. In the $\beta$-profile, one
assumes that
\be
M(r)=4\pi\rho_0r_s^3\{\ln[x+(1+x^2)^{1/2}]-x(1+x^2)^{-1/2}\}. \label{massa.beta}\ee
Assuming a NFW universal profile, the mass $M(r)$ writes as
\be
M(\leq r)=4\pi\rho_s(1+\frac{\Omega_b}{\Omega_{dm}})r_s^3m(x)  , \ee
where $m(x)$ describes the mass profile of the cluster.

There are various choices for the gas equation of state:  (a) the gas follows the Dark
Matter, $\rho_g\propto\rho_{dm}$, (b) the cluster is isothermal, $T_g=$const, (c) the gas
follows a polytropic law, $T_g\propto\rho_g^{\gamma-1}$.
In the following we will consider specifically the cases (b) and (c), and we will solve
the hydrostatic equilibrium equation for the IC gas density with and without the presence
of a magnetic field  $B(r)$.

\subsection{Isothermal case}

Let us consider first the case $B=0$.\\
In the case of a NFW mass density profile, eq.(\ref{eq.idr.om}) can be written as:
 \be
\frac{k_BT_g}{\mu m_p}\rho_g(r)^{-1}\frac{d \rho_g(r)}{d r}=-\frac{GM(\leq r)}{r^2}.
 \ee
Since
\be
M(\leq r)=M_{vir}\left[\frac{m(x)}{m(c)}\right] \label{M.M_vir}
 \ee
holds, it follows that
\be
\frac{d y_g(x)}{y_g(x)}=-\frac{GM_{vir}\mu m_p}{k_BT_gr^2}\left[\frac{m(x)}{m(c)}\right]d
r.
 \ee
After integration we find
\begin{eqnarray}
\int_0^x\frac{d y_g(x)}{y_g(x)}&=&-3\frac{T_{vir}}{T_g}\frac{c}{m(c)}\int_0^x
\frac{m(u)}{u^2}d u,\\ \int_0^x\frac{d y_g(x)}{y_g(x)}&=&-3\frac{c}{m(c)}\int_0^x
\frac{m(u)}{u^2}d u,
\end{eqnarray}
from which
\be
\rho_g(r)=\rho_g(0)e^{-3\frac{c}{m(c)}\int_0^x \frac{m(u)}{u^2}d u}. \label{NFW.profile}
 \ee
obtains.
The quantity $T_{vir}=\frac{GM_{vir}\mu m_p}{3k_Br_{vir}}$ is the virial temperature and
$\rho_g(0)$ is the central gas density.

When $B\neq0$, the hydrostatic equilibrium condition (eq.\ref{eq.idr.om}) is modified
since an additional magnetic pressure term exists:
\be
\frac{\partial p_g(r,B)}{\partial r}+\frac{\partial p_B(r,B)}{\partial r}=-\frac{GM(\leq
r)}{r^2}\rho_g(r,B),\label{eq.idr.iso.B}
\ee
where
\be
B(r)\propto B_*(\mu G)\rho_g(r,B)^{\alpha}\label{B(r)}\ee
and
\be
p_B\propto B^2. \ee
The HE condition in eq.(\ref{eq.idr.iso.B}) can be written as:
\begin{eqnarray}
\rho_g(r,B)^{-1}C_1\frac{\partial\rho_g(r,B)}{\partial
r}&+&\frac{C_2^2}{8\pi}\rho_g(r,B)^{-1} \nonumber\\
\cdot\frac{\partial\rho_g(r,B)^{2\alpha}}{\partial r}&=&-\frac{GM(\leq r)}{r^2}
\end{eqnarray}
and
\begin{eqnarray}
\rho_g(r,B)^{-1}C_1\frac{\partial\rho_g(r,B)}{\partial
r}+\frac{C_2^2}{8\pi}&&\rho_g(r,B)^{-1}
\left(\frac{2\alpha}{2\alpha-1}\right)\cdot\nonumber\\
\cdot\rho_g(r,B)\frac{\partial\rho_g(r,B)^{2\alpha-1}}{\partial r}=&-&\frac{GM(\leq
r)}{r^2},
\end{eqnarray}
where $C_1=p_g(0,B)/\rho_g(0,B)\propto k_BT_g$ and
$C_2=B_*(10^4\bar{\rho}_g(z=0))^{-\alpha}$.
It follows that
\begin{eqnarray}
y_g(x,B)^{-1}\frac{\partial y_g(x,B)}{\partial r}+ \nonumber
\end{eqnarray}
\be
+\left(\frac{2\alpha}{2\alpha-1}\right)C_1^{-1}\frac{C_2^2}{8\pi} \frac{\partial
y_g(r,B)^{2\alpha-1}}{\partial r}=-C_1^{-1}\frac{GM(\leq r)}{r^2}.\ee
After some algebra, the following equation
\begin{eqnarray}
\frac{d y_g(x,B)}{y_g(x,B)}&+&\frac{\rho_g(0,B)^{2\alpha}}{p_g(0,B)}
\left(\frac{2\alpha}{2\alpha-1}\right)\frac{C_2^2}{8\pi}d
y_g(x,B)^{2\alpha-1}=\nonumber\\
&=&-\frac{\rho_g(0,B)}{p_g(0,B)}\frac{GM(\leq r)}{r^2}dr.
\end{eqnarray}
is found, and after integration one obtains:
\begin{eqnarray}
\ln y_g(r,B)&+&\eta^{\prime}(y_g(r,B)^{2\alpha-1}-1)=\nonumber\\ &=&\frac{\ln
y_g(r,0)}{1-\frac{M_{\phi}(B)^2}{M_{vir}^2}+\frac{4\pi}{G}
\frac{r_{vir}^4}{M_{vir}^2}P_{ext}},\label{NFW.profileB}\\
\eta^{\prime}&=&\frac{C_2^2}{8\pi}\left(\frac{2\alpha}{2\alpha-1}\right)
\rho_g(0,B)^{2\alpha-1}\frac{\mu m_p}{k_BT_g(B)},
\end{eqnarray}
where $y_g(r,0)$ is defined by equation eq.(\ref{NFW.profile}). The solution of this
equation provides the radial profile of the gas density once the constants $\rho_g(0,B)$
and $T_g(B)$ are known.
The central gas density is set as
\be
\rho_g(0,B)=\rho_g(0,0),\label{norm_rho} \ee
while the temperature
\be
T_g(B)=T_g(0)\left(1-\frac{M_{\phi}(B)^2}{M_{vir}^2}+\frac{4\pi}{G}
\frac{r_{vir}^4}{M_{vir}^2}P_{ext}\right)\label{VMT} \ee
is obtained from the MVT (see Colafrancesco \& Giordano 2006a for a derivation).
For a given cluster gas mass, the density profile should be re-normalized as
\be
M_{g,vir}(B)=M_{g,vir}(B=0) \ee
from which the condition
\be
\rho_g(B,0)=\rho_g(0,0)\frac{\int dx x^2y_g(x,0)}{\int dx
x^2y_g(x,B)}.\label{norm_rhoB}\ee
 obtains.

\subsection{Polytropic case}
Let us consider first the case $B=0$.\\
Assuming
\be
p_g\propto\rho_g^{\gamma},\ee
eq.(\ref{eq.idr.om}) writes as:
\be
K_1\rho_g(r)^{-1}\frac{d \rho_g(r)^{\gamma}}{d r}=-\frac{GM(\leq r)}{r^2}, \ee
with $K_1=p_g(0)/\rho_g(0)^{\gamma}$. From this one obtains
 \be
K_1\frac{\gamma}{\gamma-1}\rho_g(0)^{\gamma-1}\frac{dy_g(r)^{\gamma-1}}
{dr}=-\frac{GM(\leq r)}{r^2},
 \ee
and
\be
\frac{dy_g(r)^{\gamma-1}}{dr}=-\left(\frac{\gamma-1}{\gamma}\right)\frac {GM(\leq r)\mu
m_p}{k_BT_g(0)r^2}.\ee
From eq.(\ref{M.M_vir}) and after integration between the limits $r=0$ e $r$, one finds
the profile
\begin{eqnarray}
\rho_g(r)&=&\rho_g(0)\cdot\nonumber\\
&\cdot&\left[1-3\frac{T_{vir}}{T_g(0)}\left(\frac{\gamma-1}
{\gamma}\right)\frac{c}{m(c)}\int_0^x\frac{m(u)}{u^2}du\right]^{\frac{1}{\gamma-1}},
\end{eqnarray}
with
\begin{eqnarray}
\frac{T_g(0)}{T_{vir}}&=&\frac{3}{\gamma}\frac{c}{m(c)} |S_*|^{-1}\\ &&\times
\left(\frac{m(c)}{c} +(\gamma-1)|S_*|\int_0^c \frac{m(u)}{u^{2}}du\right)\nonumber\\
\gamma&=&1.15+0.01(c-6.5),
\end{eqnarray}
and $S_* \equiv d\ln {y_{dm}(x)}/d\ln {x}|_c$.

For $B\neq0$, eq.(\ref{eq.idr.iso.B}) is substituted by the following equation:
\begin{eqnarray}
K_1\rho_g(r,B)^{-1}\frac{\partial\rho_g(r,B)^{\gamma}}{\partial
r}&+&\frac{K_2^2}{8\pi}\rho_g(r,B)^{-1}\cdot\nonumber\\
\cdot\frac{\partial\rho_g(r,B)^{2\alpha}} {\partial r}&=&-\frac{GM(\leq r)}{r^2},
\end{eqnarray}
and
\begin{eqnarray}
\left(\frac{\gamma}{\gamma-1}\right)\frac{\partial \rho_g(r,B)^{\gamma-1}}{\partial
r}+K_1^{-1}\frac{K_2^2}{8\pi}\left(\frac{2\alpha}{2\alpha-1}\right) \cdot \nonumber
\end{eqnarray}
\be
\cdot\frac{\partial \rho_g(r,B)^{2\alpha-1}}{\partial r}=-K_1^{-1}\frac{GM(\leq
r)}{r^2}.\ee
with $K_2\equiv C_2$.
Then, it follows
\begin{eqnarray}
\frac{\partial y_g(r,B)^{\gamma-1}}{\partial r}+\left(\frac{\gamma-1}{\gamma}\right)
\left(\frac{2\alpha}{2\alpha-1}\right)\frac{K_2^2}{8\pi}
\frac{\rho_g(0,B)^{2\alpha}}{p_g(0,B)} \cdot \nonumber
\end{eqnarray}
\be
\cdot\frac{\partial y_g(r,B)^{2\alpha-1}}{\partial
r}=-\left(\frac{\gamma-1}{\gamma}\right)\frac{\rho_g(0,B)}{p_g(0,B)} \frac{GM(\leq
r)}{r^2}.\ee
After integration one finds
\be
y_g^{\gamma -1}(x,B)+ \eta y_g^{2\alpha-1}(x,B)=(1+\eta)y_g^{\gamma
-1}(x,B=0),\label{Zhang} \ee
where
\be
\eta=\frac{p_B(r=0,B)}{p_g(r=0,B)}\frac{2\alpha}{2\alpha-1}\frac{\gamma-1}{\gamma}. \ee
To also solve numerically eq.(\ref{Zhang}) we have to know the quantities $\gamma$,
$T_g(0,B)$ and $\rho_g(0,B)$. The quantity $\gamma$ is independent of the magnetic field
$\vec{B}$; the quantities $T_g(0,B)$ and $\rho_g(0,B)$ are related to the corresponding
quantities for $B=0$ by the relations:
\be
T_g(0,B)=\frac{T_g(0,0)}{1+\eta},
\ee
As a consequence, it follows that:
 \be
\rho_g(0,B)=\frac{\rho_g(0,0)}{(1+\eta)^{\frac{1}{1-\gamma}}}.
 \ee




\begin{thebibliography}{}

\bibitem{} Arnaud, M. 2005, in 'Background Microwave Radiation and Intracluster
Cosmology', Eds. F. Melchiorri and Y. Rephaeli, (IOS Press, The Netherlands, and
Societ\'a Italiana di Fisica, Bologna, Italy, 2005) p.77 (astro-ph/0508159)
\bibitem{} Arnaud, M. \& Evrard, A. 1999, MNRAS, 305, 631
\bibitem{} Arnaud, M., Pointecouteau, E. and Pratt, G. W. 2005, A\&A, 441, 893
\bibitem{} Carilli, C.L. \& Taylor, G.B. 2002, ARA\&A, 40, 319
\bibitem{} Cassano, R., Brunetti, G. \& Setti, G. 2006, MNRAS, 369, 1577
\bibitem{} Colafrancesco, S., Marchegiani, P. and Perola, G.C. 2005, A\&A, 443, 1
\bibitem{} Colafrancesco, S., Dar, A. \& De Rujula, A. 2004, A\&A, 413, 441
\bibitem{} Colafrancesco, S. \& Giordano, F. 2006a, A\&A, 454, L131
\bibitem{} Colafrancesco, S. \& Giordano, F. 2006b, A\&A, submitted
\bibitem{} Colafrancesco, S., Profumo, S. and Ullio, P. 2006, A\&A, 455, 21
(astro-ph/0507575)
\bibitem{} Dolag, K, Bartelmann, M. and Lesch, H., 2002, A\&A, 387, 383
\bibitem{} Donahue et al. 2006, ApJ, 643, 730
\bibitem{} Eke, V. et al. 1996, MNRAS, 282, 263
\bibitem{} Evrard, A.G. \& Henry, P. 1991, ApJ, 383, 95
\bibitem{} Evrard, A.G., Metzler, C.A. \& Navarro, J.F. 1996, ApJ, 469, 494
\bibitem{} Fabian, A.C. 2005, Phil. Trans. Roy. Soc. Lond., A363, 725 (astro-ph/0407484)
\bibitem{} Fang, T. \& Bryan, G.L. 2001, ApJ, 561, L31
\bibitem{} Fujita, Y., Takizawa, M. \& Sarazin, C.L. 2003, ApJ, 584, 190
\bibitem{} Giovannini, M. 2004, International Journal of Modern Physics D,
Vol. 13, Issue 03, pp. 391-502 (astro-ph/0312614)
\bibitem{} Kaiser, N. 1991, ApJ, 383, 104
\bibitem{} Komatsu, E. and Seljak, U. 2001, MNRAS, 327, 1353
\bibitem{} Makino, N., Sasaki, S. \& Suto, Y. 1998, ApJ, 497, 555
\bibitem{} Markevitch, M. 1998,  ApJ, 504, 27
\bibitem{} Mohr, J. et al. 1999, ApJ, 517, 627
\bibitem{} Mushotzky, R.F. 2003, astro-ph/0311105
\bibitem{} Navarro, J., Frenk, C. \& White, S.D.M. 1997, ApJ, 490, 493
\bibitem{} Piffaretti, R., Jetzer, P., Kaastra, J. and Tamura, T. 2005, A\&A, 433, 101
\bibitem{} Ponman, T. et al. 1999, Nature, 397, 135
\bibitem{} Ponman, T. et al. 2003, MNRAS, 343, 331
\bibitem{} Pointecouteau, E., Arnaud, M. \& Pratt, G. W. 2005, A\&A, 435, 1
\bibitem{} Pratt, G. and Arnaud, M. 2003, A\&A, 408, 1
\bibitem{} Pratt, G. W., Arnaud, M. and Pointecouteau, E. 2006, A\&A, 446, 429
\bibitem{} Rasia, E. et al. 2006, MNRAS, 369, 2013
\bibitem{} Ryu, D. et al. 2003, ApJ, 593, 599
\bibitem{} Sarazin, C.L. 1988, 'X-Ray Emission from Clusters of Galaxies' (Cambridge
University Press: Cambridge)
\bibitem{} Seljak, U. 2000, MNRAS, 318, 203
\bibitem{} Voit, M. and Bryan, G.L. 2001, Nature, 414, 425
\bibitem{} Voit, M. et al. 2002, ApJ, 576, 601
\bibitem{} Voit, M. 2005, AdSpR, 36, 701
\bibitem{} Wu, X.-P. and Xue, Y.-J. 2002, ApJ, 572, 19
\bibitem{} Zhang, P. 2003, MNRAS, 348, 1348

\end{thebibliography}
\end{document}